\documentclass[traditabstract]{aa}  
\newcommand{\micron}{$\mu$m}

\newcommand{\cii}{[\ion{C}{ii}]}

\newcommand{\kms}{km\,s$^{-1}$}

\usepackage{graphicx}
\usepackage{txfonts}

\begin{document}

   \title{Probing the 2D temperature structure of protoplanetary disks with {\it Herschel} observations of high-$J$ CO lines}
%		\subtitle{Measured with high-$J$ CO lines with {\it Herschel}}
   \author{
	D.\ Fedele\inst{\ref{inst_inaf}, \ref{inst_mpe}},
  E.F. \ van Dishoeck\inst{\ref{inst_mpe}, \ref{inst_leiden}},
	M. \ Kama\inst{\ref{inst_leiden}},
	S. \ Bruderer\inst{\ref{inst_mpe}}, 
	M.R. \ Hogerheijde\inst{\ref{inst_leiden}}
}
      
    \institute{
INAF-Osservatorio Astrofisico di Arcetri, L.go E. Fermi 5, I-50125 Firenze, Italy\label{inst_inaf},\\
\and
Max Planck Institut f\"{u}r Extraterrestrische Physik, Giessenbachstrasse 1, 85748 Garching, Germany\label{inst_mpe},\\
\and
Leiden Observatory, Leiden University, P.O. Box 9513, NL-2300 RA, Leiden, The Netherlands\label{inst_leiden}
}

\authorrunning{Fedele et al.}
\offprints{Davide Fedele,\\ \email{fedele@mpe.mpg.de}}

\abstract{
The gas temperature structure of protoplanetary disks is a key ingredient for interpreting various disk observations and for quantifying the subsequent evolution of these systems. The comparison of low- and mid-$J$ CO rotational lines is a powerful tool to assess the temperature gradient in the warm molecular layer of disks. Spectrally resolved high-$J$ ($J_{\rm u} > 14$) CO lines probe intermediate distances and heights from the star that are not sampled by (sub-)millimeter CO spectroscopy.  This paper presents new {\it Herschel}/HIFI and archival PACS observations of $^{12}$CO, $^{13}$CO and \cii \ emission in 
4 Herbig AeBe (HD 100546, HD 97048, IRS 48, HD 163296) and 3 T Tauri (AS 205, S CrA, TW Hya) disks. In the case of the T Tauri systems AS 205 and S CrA, the CO emission has a single-peaked profile, likely due to a slow wind. For all other systems, the {\it Herschel} CO spectra are consistent with pure disk emission and the
spectrally-resolved lines (HIFI) and the CO rotational ladder (PACS) are analyzed simultaneously assuming power-law temperature and column density
profiles, using the velocity profile to locate the emission in the disk.
The temperature profile varies substantially from disk to disk. In particular, $T_{\rm gas}$ in the disk surface layers can differ by
up to an order of magnitude among the 4 Herbig AeBe systems with HD 100546 being the hottest and HD 163296 the coldest disk of the sample. 
Clear evidence of a warm disk layer where $T_{\rm gas} > T_{\rm dust}$ is found in all the Herbig Ae disks.
The observed CO fluxes and line profiles are compared to predictions of physical-chemical models. 
The primary parameters affecting the disk temperature structure are the flaring angle, the gas-to-dust mass ratio the scale height and the dust settling. 
}
 

   \maketitle

\section{Introduction}
A key physical parameter of protoplanetary disks is the gas temperature, $T_{\rm gas}$. Inside a disk $T_{\rm gas}$ controls the dynamics of the gas 
by setting the sound speed and, through that, also the disk photoevaporation. At the same time $T_{\rm gas}$ governs the chemical composition by regulating the reaction rates 
between different species. 
Disks are characterized by a strong temperature gradient both in the radial and vertical directions. For this reason, multiple transitions, tracing different
vertical layers and different orbital radii, have to be observed to derive $T_{\rm gas}$ in disks. An ideal disk `thermometer' is the CO rotational ladder. 
Low-$J$ ($J_{\rm u} < 6$) CO rotational lines are routinely observed from the ground in protoplanetary disks since 
the late 90s \citep[e.g.,][]{Koerner95, Mannings97, Dutrey98, vanzadelhoff01}. These lines probe mostly the cold gas in the 
outer disk ($r > 100\,$au). Recent observations of disks with {\it Herschel}/PACS carried out by the DIGIT \citep{Green13} and GASPS \citep{Dent13} key programs
report the detection of pure rotational high-$J$ ($J_{\rm u}$ $>$ 14) CO emission lines \citep[e.g.,][]{Sturm10, vankempen10, Meeus12, Meeus13}. These lines  
trace warm gas ($E_{\rm u}$ $\ge$ 300\,K) located in intermediate layers between the disk surface and the midplane at
intermediate distances from the star ($10-50\,$au) as predicted by thermo-chemical models of UV irradiated disks 
\citep[e.g.,][]{Jonkheid07, Gorti08, Woitke09, Kamp10, Bruderer12}.

\smallskip
\noindent
The detections of the CO high-$J$ lines allow us, for the first time, to estimate the gas temperature  in this region of the disk. However, the PACS 
spectra (resolving power $R = \lambda/\Delta \lambda \sim 10^3$) presented by \citet{Meeus13} are spectrally and spatially unresolved, so that their emitting region (hence, the radial 
distribution of the gas) can only be inferred indirectly from the modeling of line fluxes \citep[e.g.][]{Bruderer12}.  
The only way to overcome the lack of spatial resolution at high THz frequencies, and to determine the warm gas
distribution within disks is with high-resolution spectroscopy with HIFI ($R = 10^6 - 10^7$), where Kepler's law can 
be used to associate a velocity bin with a radial location in the disk.

\smallskip
\noindent
The {\it Herschel} PACS and HIFI CO spectra of the Herbig Ae system HD 100546 have been presented in \citet{Fedele13b} 
(hereafter paper I) in which the radial gas temperature gradient is estimated for the first time.  
This paper presents new {\it Herschel}/HIFI observations of CO $J=16-15$ toward: HD 97048,
AS 205, Oph-IRS 2-48 and S CrA, CO and $^{13}$CO $J=10-9$ toward TW Hya, HD 100546 and HD 163296.
{\it Herschel}/HIFI observations of \cii \ (158\,\micron) in HD 97048 and HD 100546 are reported in the Appendix.   

\begin{table*}
\caption{{\it Herschel}/HIFI observations log and line properties. }
\centering
\begin{tabular}{llllllllll}
\hline\hline
Target & RA      & DEC     & Obsid  & v$_{\rm LSR}$ & FWHM & rms & dv &Int. Intensity & Int. Flux              \\
       & (J2000) & (J2000) & 13422- & [\kms]        & [\kms] & [K] & [\kms] & [K~\kms]     & [10$^{-17}$\,W\,m$^{-2}$]\\
\hline
\\
\multicolumn{10}{c}{CO $J=16-15$, E$_{\rm u}$ = 751.8\,K, 1841.345\,GHz, $\eta_{\rm mb}$ = 0.57 (H), 0.60 (V), HPBW = 11.1\,$\arcsec$}\\
\\
HD 97048          & 11:08:03.32 & -77:39:17.5 & 50973 & 4.9 & 7.4  & 0.04 & 0.08 & 1.42 $\pm$ 0.05  & 2.97 $\pm$ 0.10 \\
HD 100546         & 11:33:25.44 & -70:11:41.2 & 47519 & 5.3 & 8.2  & 0.06 & 0.08 & 2.89 $\pm$ 0.08  & 6.06 $\pm$ 0.17 \\
AS 205            & 16:11:31.40 & -18:38:24.5 & 51072 & 4.8 & 3.7  & 0.04 & 0.08 & 0.76 $\pm$ 0.06  & 1.59 $\pm$ 0.12 \\
Oph-IRS 48        & 16:27:37.19 & -24:30:35.0 & 51070 & 4.8 & 12.0 & 0.04 & 0.08 & 0.40 $\pm$ 0.04  & 0.84 $\pm$ 0.08 \\
S CrA             & 19:01:08.60 & -36:57:20.0 & 53691 & 6.4 & 2.6  & 0.04 & 0.08 & 1.35 $\pm$ 0.07  & 2.83 $\pm$ 0.15 \\
\\
\multicolumn{10}{c}{CO $J=10-9$, E$_{\rm u}$ = 304.2\,K, 1151.985\,GHz, $\eta_{\rm mb}$ = 0.59 (H), 0.59 (V), HPBW = 19.5\,$\arcsec$} \\
\\
TW Hya            & 11:01:51.91 & -34:42:17.0 & 10733 & 2.9 & 1.3 & 0.06 & 0.13 & 0.22 $\pm$ 0.04 & 0.34 $\pm$ 0.06 \\%$<$ 0.06\tablefootmark{$\star$} & $<$ 0.3\\
HD 100546         & 11:33:25.44 & -70:11:41.2 & 35779 & 5.5 & 6.4 & 0.04 & 0.13 & 3.29 $\pm$ 0.05 & 5.21 $\pm$ 0.08 \\
HD 163296         & 17:56:21.29 & -21:57:21.9 & 51440 & 5.9 & 6.5 & 0.07 & 0.13 & 0.95 $\pm$ 0.23 & 1.51 $\pm$ 0.37 \\
\\
\multicolumn{10}{c}{$^{13}$CO $J=10-9$, E$_{\rm u}$ = 290.8\,K, 1101.349\,GHz , $\eta_{\rm mb}$ = 0.59 (H), 0.59 (V), HPBW = 19.5\,$\arcsec$}\\
\\
TW Hya            & 11:01:51.91 & -34:42:17.0 & 01585 & 2.7 & 1.2 & 0.006 & 0.14 & 0.02 $\pm$ 0.003 & 0.028 $\pm$ 0.004\\
HD 100546         & 11:33:25.44 & -70:11:41.2 & 56430 & 5.5 & 7.8 & 0.007 & 0.14 & 0.64 $\pm$ 0.03  & 0.88  $\pm$ 0.06 \\
HD 163296         & 17:56:21.29 & -21:57:21.9 & 53595 & 5.9 & 9.5 & 0.006 & 0.14 & 0.14 $\pm$ 0.01  & 0.19  $\pm$ 0.01\\
\\
\multicolumn{10}{c}{ \cii \ $^2P^{\, 0}_{3/2} - ^2P^{\, 0}_{1/2}$, E$_{\rm u}$ = 91.3\,K, 1900.537\,GHz, $\eta_{\rm mb}$ = 0.57 (H), 0.60 (V), HPBW = 11.1\,$\arcsec$}\\
\\
HD 97048          & 11:08:03.32 & -77:39:17.5 & 52194 & 4.90 & 2.75 & 0.10 & 0.08 & 4.09 $\pm$ 0.09  & 9.44 $\pm$ 0.21 \\
HD 100546         & 11:33:25.44 & -70:11:41.2 & 47518 & 5.75 & 3.70 & 0.09 & 0.08 & 8.50 $\pm$ 0.15  & 19.6 $\pm$ 0.34 \\
\hline
\end{tabular}\label{tab:log}
\tablefoot{
The rms is measured at the original spectral resolution, channel width given in column ``dv".
%\tablefoottext{$\star$}{The rms is measured at the original spectral resolution, channel width given in column ``dv"}
}
\end{table*} 

\section{Observations and data reduction}
The sample selection is based on the PACS detection of high-$J$ CO and \cii \ emission \citep{Meeus13, Fedele13a}. 
The observations log is reported in Table~\ref{tab:log}. Most of the data are taken from program ID OT2$\_$DFedele$\_$1 
(PI: D. Fedele). 
The HIFI $^{12}$CO and $^{13}$CO $J=10-9$ spectra are obtained from the WISH key program (PI: E.F. van Dishoeck) for TW Hya, and from programs 
OT1$\_$mhogerhe$\_$1 (HD 100546 $^{12}$CO $J=10-9$, PI: M. Hogerheijde), OT2$\_$mhogerhe$\_$2 (HD 100546 and HD 163296 $^{13}$CO $J=10-9$, PI: M. Hogerheijde) 
and OT1$\_$lpodio$\_$1 (HD 163296 $^{12}$CO $J=10-9$, PI: L. Podio).

\smallskip
\noindent
The CO observations were executed in dual beam switch fast chopping mode with the Wide-Band Spectrometer 
(WBS) and the High Resolution Spectrometer (HRS) simultaneously. The spectral resolution is set to 1.1\,MHz 
for WBS and 0.25\,MHz for HRS for both polarizations. The \cii \ observations were carried out in 'load chop' where 
an internal calibration source is used in combination with an off-source calibration observation. This allows us 
to remove the spatially extended \cii \ emission. The beam size (HPBW) is 11\farcs1 at the observed frequency 
\citep{Roelfsema12}. In the following, we will refer only to the WBS spectra.

\smallskip
\noindent
The spectra are extracted from the level 2 data which have been processed with standard pipeline SPG v9.1.0. Standing 
waves are present in the WBS spectra. These have been removed by fitting a set of sine functions after masking 
the narrow spectral features (CO or \cii). This operation was performed with the 'fitHifiFringe' script provided with 
Hipe. The HIFI level 2 fluxes are given on the antenna temperature scale ($T_{\rm A}^*$). These are converted to main
beam temperature, $T_{\rm mb} = T_{\rm A}^* \times \eta_{\rm l}/\eta_{\rm mb}$, with $\eta_{\rm l}$ the forward efficiency 
and $\eta_{\rm mb}$ the beam efficiency (Table~\ref{tab:log}). No major differences are present between the H and V 
polarizations and the two spectra are averaged together after applying the efficiency corrections and removing the continuum. 
High-$J$ CO lines are not contaminated by the cold cloud contribution that plagues single dish low-$J$ CO lines. 

\smallskip
\noindent
The reduction of the archival PACS data analysed here is described in \citet{Meeus13} and \citet{Fedele13a}.

\begin{figure*}[!t]
\includegraphics[width=18cm]{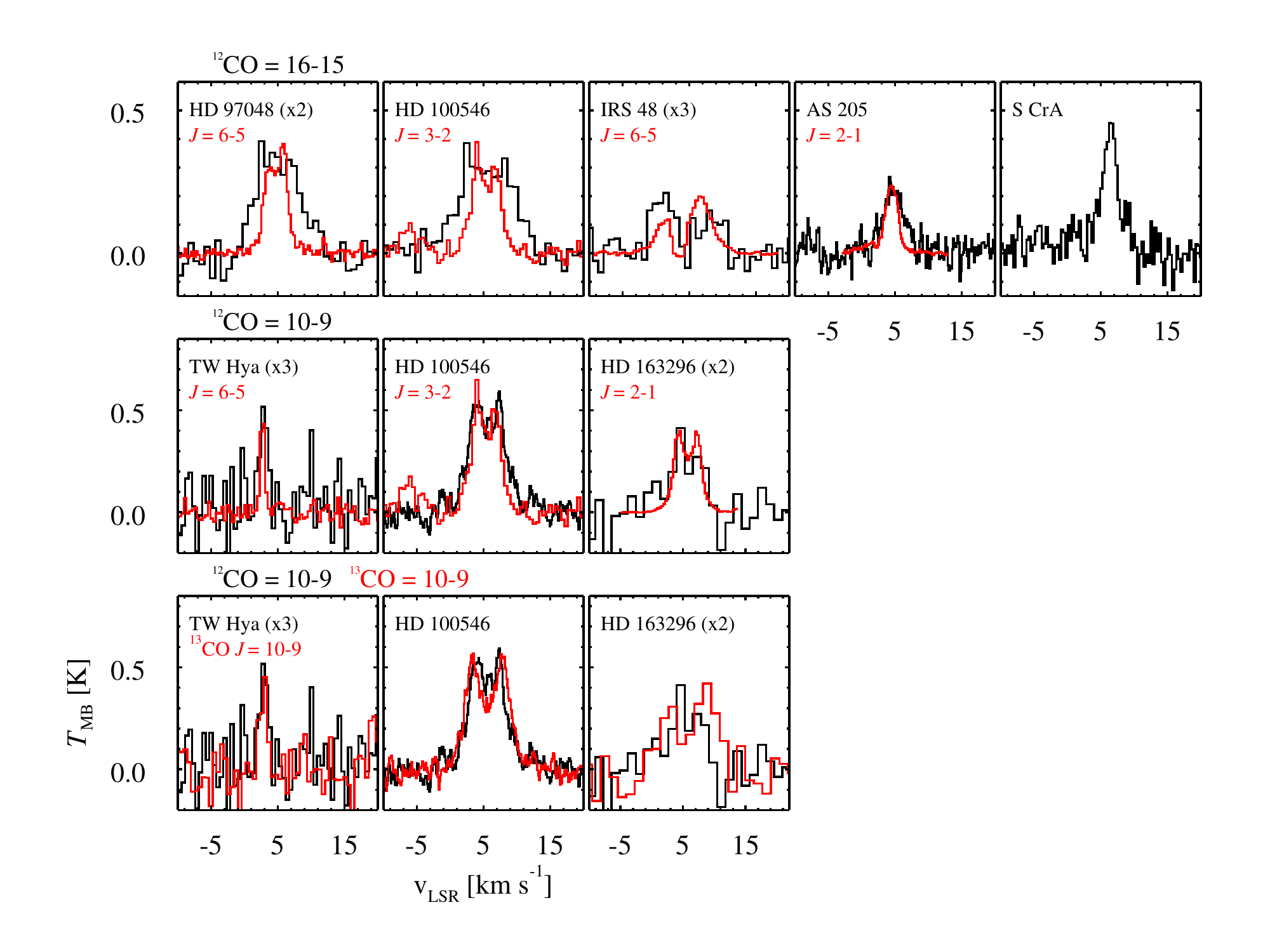}
\caption{HIFI/WBS spectra of CO $J$=16-15 (top), $J=10-9$ (middle and bottom) and $^{13}$CO $J=10-9$ (bottom). The spectra of low-$J$ lines are overlaid in red (scaled for comparison) in top and middle rows. 
For clarity, some spectra are rebinned to lower resolution: for CO $J=16-15$ dv = 0.8\,\kms \ in HD 97048, HD 100546 and IRS 48 and dv = 0.32\,\kms in S CrA; for 
CO $J=10-9$ dv = 0.52\,\kms \ in TW Hya and dv = 1.3\,\kms \ in HD 163296; for $^{13}$CO $J=10-9$ dv = 0.56\,\kms \ in TW Hya and dv = 1.4\,\kms in HD 163296. The remaining spectra are shown at their native resolution.}
\label{fig:spectra}
\end{figure*}

\begin{figure*}
\centering
\includegraphics[width=12cm]{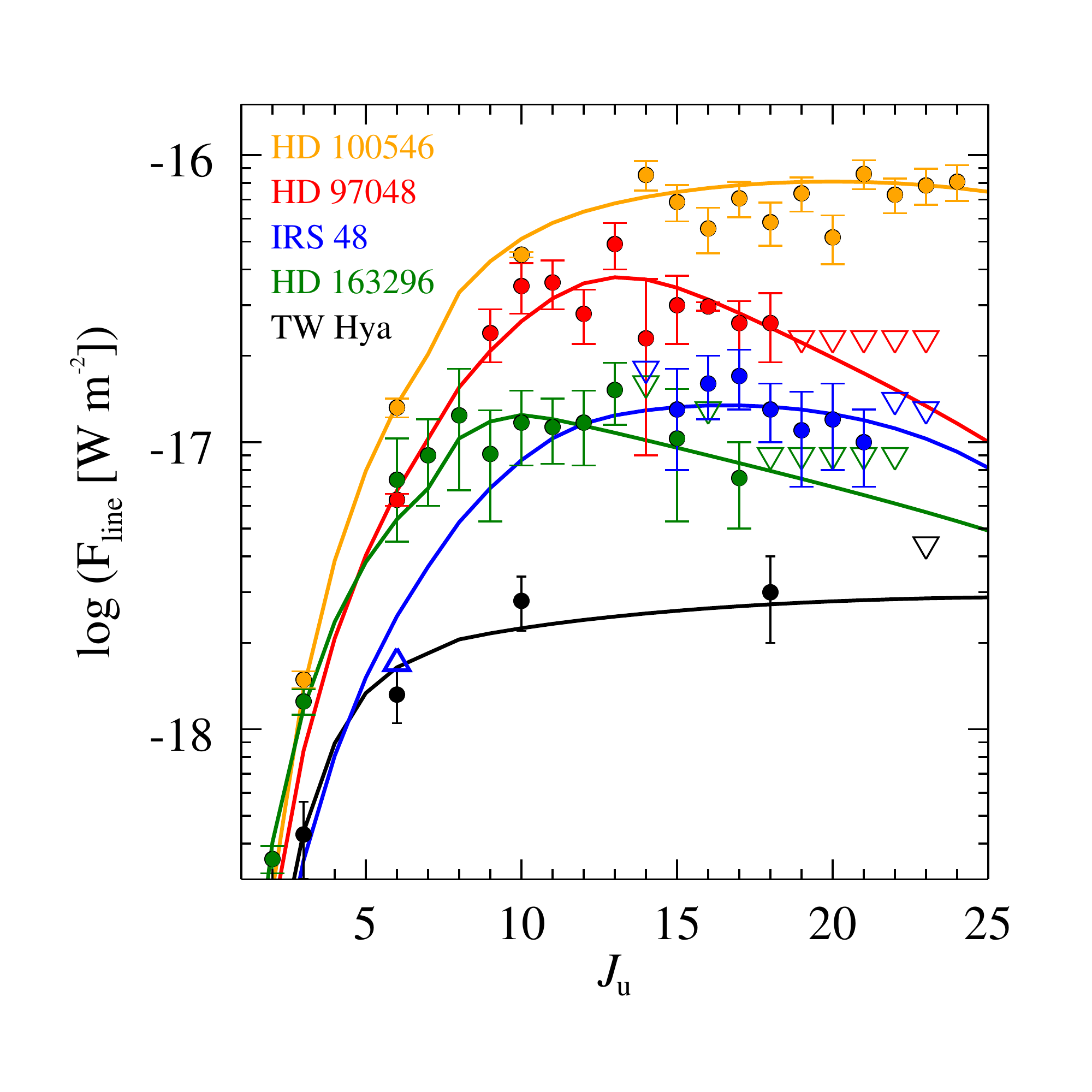}
\includegraphics[width=18cm]{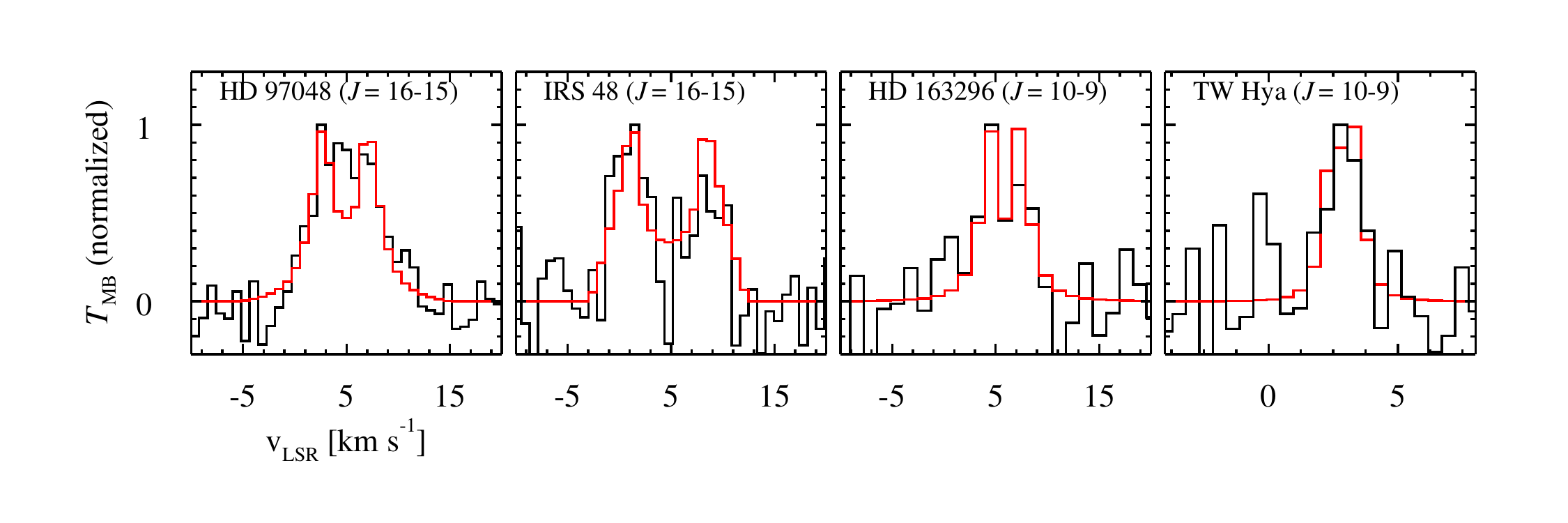}
\caption{(top) CO rotational ladder. Detections are shown as filled circles and upper limits as open triangles. The solid 
lines represent the best-fit power-law model. (bottom) Best-fit model profile of the CO $J=16-15$ (HD 97048 and IRS 48) and CO $J=10-9$ (HD 163296 and TW Hya).
Spectral resolution as in Figure~\ref{fig:spectra}.}
\label{fig:coladder}
\end{figure*}

\section{Results}\label{sec:results}
The HIFI/WBS CO spectra are presented in Figure~\ref{fig:spectra} and the line parameters are given in Table~\ref{tab:log}. 
The \cii \ spectra and analysis is presented in the Appendix.
The integrated line flux (W\,m$^{-2}$) is computed from the integrated intensity

\begin{equation}
\int{T_{\rm mb}dV}  = 2 \ k \ \Big(\frac{\nu}{c} \Big)^3 \ \pi \ \Bigg(\frac{\rm HPBW}{2 \ \sqrt{\rm ln (2)}}\Bigg)^2 \ \int{T_{\rm mb}dV} \ \ \ {\rm [K\,km\,s^{-1}]}
\end{equation}
%[K\,\kms]

\noindent
with $k$ (Boltzmann constant, W\,s\,K$^{-1}$), $\nu$ the frequency (Hz), $c$ the light speed (m\,s$^{-1}$) and 
HPBW the beam (radians). Note that because of the updated values of the beam efficiency (Table~\ref{tab:log}), 
the line intensities presented here for HD 100546 are slightly different ($\sim 3\,\%$) from the values given in paper I. 

\smallskip
\noindent
The top panel of Figure~\ref{fig:spectra} shows the WBS spectra of the CO $J=16-15$ lines toward HD 97048, HD 100546, IRS 48, 
AS 205 and S CrA. The line is clearly detected above 5\,$\sigma$ (Table~\ref{tab:log}) in all sources. The velocity profile and width are 
different among the five sources. The emission is broad ($\Delta v > 5\,$\kms) toward the three HAeBe stars HD 97048, HD 100546 
and IRS 48: a double-peak profile is clearly visible in HD 100546 and IRS 48, while the CO velocity profile is top-flat in HD 97048. 
In all three cases, the WBS spectra of the CO $J=16-15$ transition are consistent with a Keplerian velocity field of the gas in the disk. 
In the case of the two T Tauri systems, AS 205 and S CrA, the CO $J=16-15$ 
emission is narrow ($\Delta v < 5\,$\kms) and single-peaked. There is no evidence of Keplerian rotation. 
For comparison, the profiles of the low$-J$ CO lines are also shown when available. Spectra (either APEX or ALMA) are from \citet{Panic10, Bruderer14, Salyk14} 
and Kama et al. (in prep.). 
In all cases, the low and high-$J$ CO lines are centered at the same v$_{\rm LSR}$ with the CO $J=16-15$ line broader than the low$-J$ one. Note that part of
the asymmetric profile of the $J=6-5$ line toward IRS 48 is due to extinction from the foreground cloud \citep{Bruderer14}. 
In the case of AS 205 both CO lines are centered at v$_{\rm LSR} = 4.8$\,\kms \ compared to the cloud velocity of v$_{\rm LSR} = 3$\,\kms.  Thus we conclude that 
the high-$J$ CO emission in AS 205  arises from a slow wind/outflow similarly to the low-$J$ CO \citep{Salyk14} and to the ro-vibrational \citep{Pontoppidan11} emission.  

\smallskip
\noindent
The narrow, single-peak, profile of the CO $J=16-15$ emission in S CrA suggests also a contribution from a slow wind associated to the system. The CO ro-vibrational lines toward both stellar components in this binary are broad and single-peaked, similar to AS 205 \citep{Bast11, Brown13}.    

\smallskip
\noindent
Figure~\ref{fig:spectra} also shows the HIFI/WBS spectra of CO $J=10-9$ compared to the low$-J$ transitions (middle) and
to the $^{13}$CO (bottom) for TW Hya, HD 100546 and HD 163296. Both the $^{12}$CO and $^{13}$CO lines are clearly detected. The 
lines are broad and double-peaked toward HD 100546 and HD 163296 and narrow and single-peaked toward TW Hya. In all 3 cases, the lines are centered 
on the system velocity and the narrow line profile of TW Hya is consistent with the disk being almost face-on to the plane of the sky. 
As noted in paper I, the line width is narrower for lower $J$ transitions in HD 100546. In the case of HD 163296 instead, the width of the $J=2-1$ 
(ALMA science verification data) and $J=10-9$ are similar. Note that the $^{12}$CO and $^{13}$CO $J=10-9$ profiles appear asymmetric toward HD 163296, 
however the flux difference between the two peaks is within the noise level of the spectrum. 

\smallskip
\noindent
Interestingly, the $^{13}$CO line toward HD 100546 and HD 163296 is slightly broader than the $^{12}$CO one (Figure~\ref{fig:spectra}, bottom row). 
This implies that the line emitting 
region of $^{13}$CO extends to higher velocity regions, i.e. closer to the star. 
Another prominent difference is the central peak detected in the $^{12}$CO $J=10-9$ line in the HD 100546 spectrum which is not visible in the 
$^{13}$CO spectrum. The differences in the velocity profiles are likely due to optical depth effects: the $^{12}$CO line becomes optically 
thick at lower column densities, higher up in the atmosphere, than the $^{13}$CO line. This is discussed further in section~\ref{sec:dali}.

\begin{figure}[!h]
\centering
\includegraphics[width=9cm]{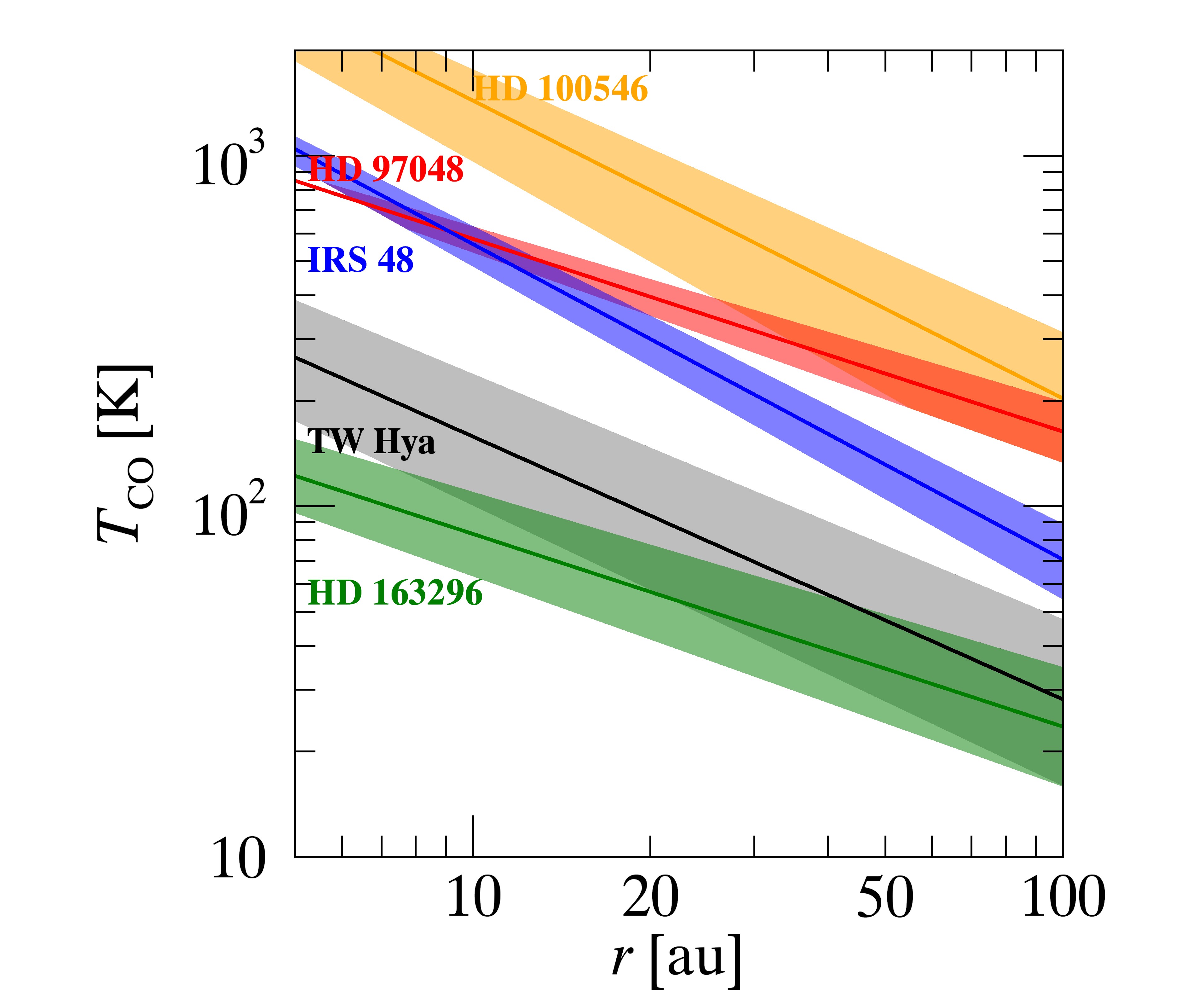}
\caption{CO temperature profile in the inner 100\,au based on the best-fit power-law models (Table~\ref{tab:param}).}
\label{fig:tgas}
\end{figure}

\begin{table*}
\caption{Best-fit power-law model parameters}
\centering
\begin{tabular}{llllll}
\hline\hline
 &   TW Hya & HD 97048 & HD 100546\tablefootmark{$\dagger$} & HD 163296 & IRS 48 \\
\hline
$M_{\star}$    (M$_{\odot}$) & 0.6           & 2.5    & 2.5   & 2.4       & 2.0 \\
$d$            (pc)                     & 51            & 180    & 97      & 120      & 120 \\
$i$            ($^{\circ}$)           & 7               & 43     & 42       & 44        & 50 \\
$r_{\rm i}$    (au)                  & 0.1            & 11     & 13       & 0.1       & 20 \\
$r_{\rm out}$  (au)                  & 200           & 400 & 400& 600 & 200\\ 
& & & & & \\
$T_{\rm i}$    (K)                  & $3000-5000$   & $500-600$ & 750-1450 & $1000 - 1100$       & $250-350$\\
$q$                                       & $0.7-0.8$        & $0.5-0.6$ & $0.75 - 0.95$ &  $0.5 - 0.6$         & $0.85-0.95$ \\
$N_{\rm i}$    (cm$^{-2}$)   & $1-5 \times 10^{18}$  & $10^{18}-10^{19}$ & $2-8 \times 10^{-17}$ &$10^{19} - 10^{20}$ & $10^{18} - 10^{19}$\\
$p$                                       & $0.85-0.95$           & $0.45-0.55$ & 0.6-1.2 & $0.3-0.35$          & $0.55 - 0.65$\\
\hline
\end{tabular}\label{tab:param}
\tablefoot{
\tablefoottext{$\dagger$}{From paper I.}
Best fit parameters are found by $\chi^2$ minimization. Ranges represent the 1-$\sigma$ interval. 
}
\end{table*}

\section{Analysis}
The CO rotational ladder is a powerful tool to assess the temperature structure of protoplanetary disks 
\citep{Bruderer12, Fedele13b, vanderwiel14}. The low-$J$ lines are optically thick and the lines becomes 
optically thin only at $J_u \gtrsim 14$ \citep{Bruderer12}. The advantage of the high-$J$ CO rotational 
lines is that they trace intermediate distances from the star (a few tens of au) and heights ($z/r \sim 0.1-0.4$) 
above the disk midplane. The high-$J$ CO transitions are complimentary to the low$-J$ CO transitions observed at millimeter 
observations which trace the colder outer disk. 
Thus the flux and velocity profile of the high-$J$ CO lines are crucial to measure the gas temperature in the disk atmosphere. 
The rotational ladder (Table~\ref{tab:coladder}) is compiled combining low-$J$ lines ($J_u < 6$) 
from ground based observations, mid-$J$ ($J_u = 7-12$) from SPIRE \citep{vanderwiel14} and high-$J$ 
($J_u > 14$) lines from {\it Herschel}/PACS \citep{Meeus13}. In some cases, the CO line fluxes measured with SPIRE
are contaminated by extended emission (cloud) as noted by \citet{vanderwiel14}. The lines which are affected are excluded from the fit. 

\smallskip
\noindent
The CO rotational ladders for the 5 disks are shown in Figure~\ref{fig:coladder}. The shape of the ladder varies from object to object. Note in particular
the difference between HD 100546 and HD 163296: in the first case the CO line flux increases with $J$ and then it remains almost constant while in the
case of HD 163296, the rotational ladder turns over at $J_{\rm u} \sim 10-15$.

\smallskip
\noindent
The analysis of HD 100546 is presented in paper I. The sources AS 205 and S CrA
are excluded from this analysis because the CO line fluxes and velocity profiles  are dominated by an outflow/jet.

\subsection{Power-law model}
Following the method described in paper I the CO rotational ladder and the line velocity profiles ($J=16-15$ for 
HD 97048 and IRS 48, $J=10-9$ for TW Hya and HD 163296) are fitted simultaneously using a power-law profile for the kinetic gas temperature 
(under the assumption that the CO excitation temperature corresponds to the kinetic temperature, which is valid given the high densities of the emitting regions) and column density:

\begin{eqnarray}
T(r)  =  T_i \ \bigg ( \frac{r}{r_{\rm i}} \bigg)^{-q}\\
N(r)  =  N_i \ \bigg( \frac{r}{r_{\rm i}} \bigg)^{-p}
\end{eqnarray}

\noindent
where $T_{\rm i}$ and $N_{\rm i}$ are the values at the inner radius $r_{\rm i}$ of the disk (fixed, Table~\ref{tab:param}). 
The stellar and disk parameters are taken from literature. In particular r$_{\rm i} = 0.1\,$au for TW Hya \citep{Pontoppidan08}, 13\,au for 
HD 100546 \citep{vanderplas09, Brittain09, Fedele15}, 11\,au for HD 97048 \citep{vanderplas09} and 20\,au for IRS 48 \citep{Bruderer14}.
The power-law is truncated at the outer disk radius, $r_{\rm out}$ (fixed, Table~\ref{tab:param}, the choice of the outer radius does not matter as long as $r_{\rm out} > 100\,$au). 
The free parameters of the model are: 
$T_{\rm i}, q, N_{\rm i}, p$. A grid of models is created for each disk varying the 4 parameters 
in the ranges: $T_{\rm i} = 300 - 1500\,$K, $N_{\rm i} = 10^{17} - 10^{22}\,$cm$^{-2}$, $q, p = 0.5 - 1.5$. 
The spectra are spatially convolved with the telescope beam, represented here by a Gaussian profile.
Further details about the model and the fitting procedure are given in paper I.
The best-fit parameters are found by minimizing the $\chi^2$ between observations and model. The final $\chi^2$ 
is given by the sum of the individual $\chi^2$ of the CO rotational ladder and of the line profiles (one for each line), including their width and peak separation.   
The best-fit model parameters are listed in Table~\ref{tab:param} and the best-fit models are overlaid on the data in Figure~\ref{fig:coladder}. 

\smallskip
\noindent
The derived temperature profiles (labeled $T_{\rm CO}$) are plotted in Figure~\ref{fig:tgas}. Evidence of warm disk temperature ($T > 100\,$K) is found for HD 97048, HD 100546 and IRS 48.
The coldest disks are TW Hya and HD 163296 both having $T < 100\,$K outward of 20\,au.

  \begin{table*}
	\caption{DALI model parameters for Herbig Ae disks}
	\centering
	\begin{tabular}{lccl}
	\hline\hline
	Parameter & Value & Unit & Description \\
	\hline
	$M_{\star}$    & 2                               & [$M_{\odot}$]  & Stellar mass\\
        $T_{\rm eff}$  & 8000, \ {\bf 9000}, \  10500  &  [K]         & Stellar (blackbody) temperature       \\
	$L_{\rm bol}$  & 20                              & [$L_{\odot}$]  & Bolometric luminosity\\
 	$\Delta_{\rm gas/dust}$    & 1, \ {\bf 10} , \ 100 &             & Gas-to-dust mass ratio      \\
	PAHs                & {\bf 1}, \ 10, \ 50   & [\% (w.r.t. ISM)] & PAHs abundance        \\
	$M_{\rm disk}$ & 10$^{-4}$, \ {\bf 10$^{-3}$}, 10$^{-2}$ & [$M_{\odot}$]  & Disk mass        \\
	$\gamma$       & 0.8, \ {\bf 1}, \ 1.2          &   & Surface density power-law exponent                         \\
	$R_{\rm sub}$ & 0.31                             & [au] & Sublimation radius \\
	$R_{\rm c}$    & 50, \ {\bf 75}, \ 100          & [au] & Critical radius                              \\
	$h_{\rm c}$     & 0.1, \ {\bf 0.2}, \  0.3      & [radians]  & Scale height \\
	$R_{\rm out}$    & 200, \ {\bf 400}, \ 600      & [au]    & Disk outer radius                               \\
	$\psi$             & {\bf 0.05}, \ 0.15, \ {\bf 0.25} & & Flaring angle                            \\	
        $\chi$              & {\bf 0.2}, \ 0.5, \ 1.0   & & Degree of settling                              \\
	$f_{\rm large}$&   0.50, \ {\bf 0.85}, \ 0.999    & & Large-to-small grains mass ratio                             \\
	$a_{\rm min}$&    {\bf 0.001}, \ 0.01, \  0.05 & [\micron]     & Minimum grain size                            \\
	$L_{\rm X}$    & $10^{29}$ & [erg\,s$^{-1}$]       & X-rays luminosity \\
        $d$           & 100 & [au] & Distance \\
        $i$           & 45 & [$^{\circ}$] & Disk inclination  \\
	\hline
	\end{tabular}\label{tab:dali}
	\tablefoot{
Values in bold face indicate the representative disk models. For model description and  parameters definition see \citet{Bruderer13}.
}
	\end{table*}

\subsection{Caveats of the power-law model}
The assumption of a flat disk geometry is justified by the results of physical-chemical models that predict that most of the CO
rotational lines arise from a similar vertical layer in the disk. With this assumption however, the power-law model fails to reproduce
the core of the line profile, especially in the case of HD 97048 (Figure~\ref{fig:coladder}) where the line wings are well reproduced 
by the model while the central low velocity part of the line is not. This can be due to an optical depth effect (if the core of the line is optically thick) 
or to a geometrical effect. In this case indeed, given the disk inclination (43$^{\circ}$), the core of the line in the {\it Herschel} 
beam may be filled-in by emission coming from the back side of the disk as found for the lower $J$ CO lines from ALMA data by \citet{degregorio13}.

\smallskip
\noindent
The second major caveat of the power-law model is that the column density profile is not constrained: most of the lines
are optically thick and the CO ladder is mostly sensitive to the temperature profile of the $\tau_{\rm line}=1$ layer. Thus, even if
$N_{\rm in}$ and $p$ are free parameters of the model, these values are to be taken with caution. 

\section{Comparison to disk models}\label{sec:dali}
The goal of this section is to explain the large variation in the CO rotational ladder and velocity profiles among Herbig Ae systems 
(Figures~\ref{fig:spectra} and \ref{fig:coladder}). This study is based on the physical-chemical model \textsc{DALI} 
\citep{Bruderer12, Bruderer13}. \textsc{DALI} takes as input a density structure of the disk (taken to be a power-law with slope 
$\gamma$ and critical radius $R_{\rm c}$, with an exponential tail) and a stellar radiation field, then solves the continuum radiative transfer and 
determines the dust temperature and ultraviolet radiation field at each position in the disk. Thermal balance of the gas and chemistry are 
subsequently solved iteratively until convergence. The output includes: continuum and line emission maps,  line intensities and spectra produced via ray tracing.
Dust settling is included adopting two grain size populations, small ($a_{\rm min}$ - 1\,\micron) 
and large (1 - 1000\,\micron) following \citet{dAlessio06} and dust cross sections from \citet{Andrews11}.
The degree of settling of the large grains is controlled in DALI by the parameters $\chi$ and $f_{\rm large}$: the first defines the maximum scale 
height of the large grains with respect to the small ones (similar to $``Z_{\rm big}"$ in \citealt{dAlessio06}), while $f_{\rm large}$ determines the 
small-to-large grains mass ratio.

\begin{figure*}
\centering
\includegraphics[width=8cm]{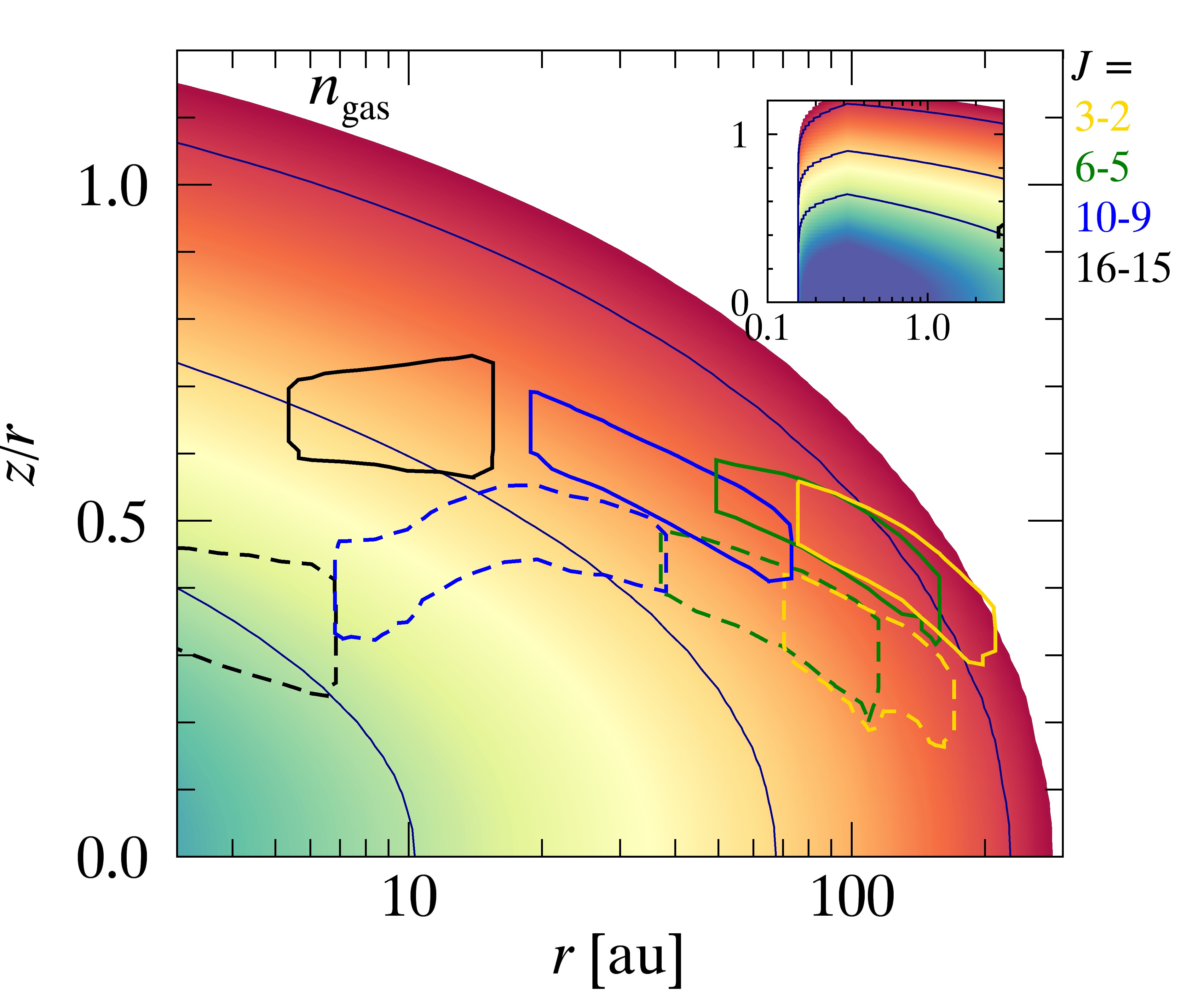}
\includegraphics[width=8cm]{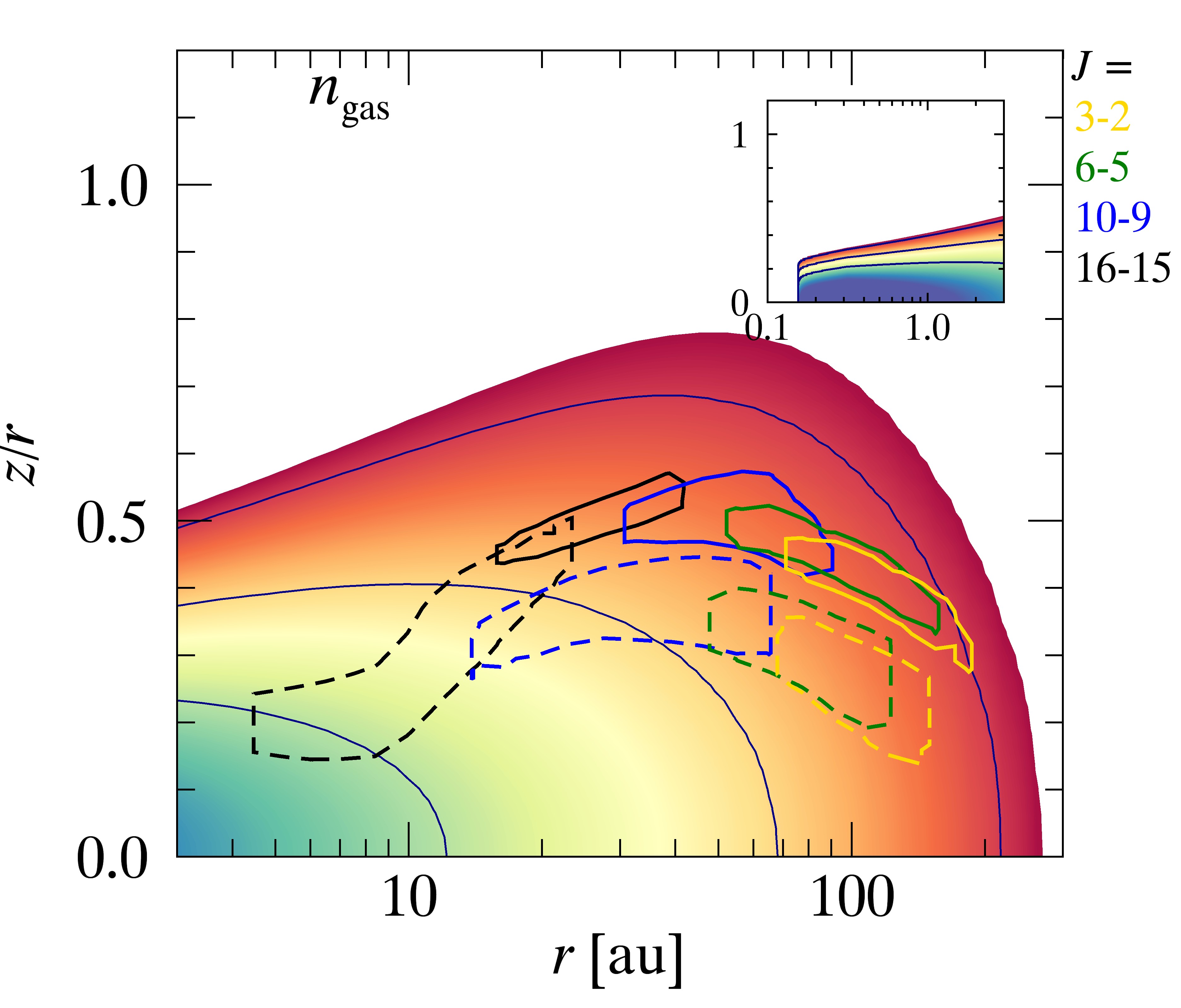}
\includegraphics[width=8cm]{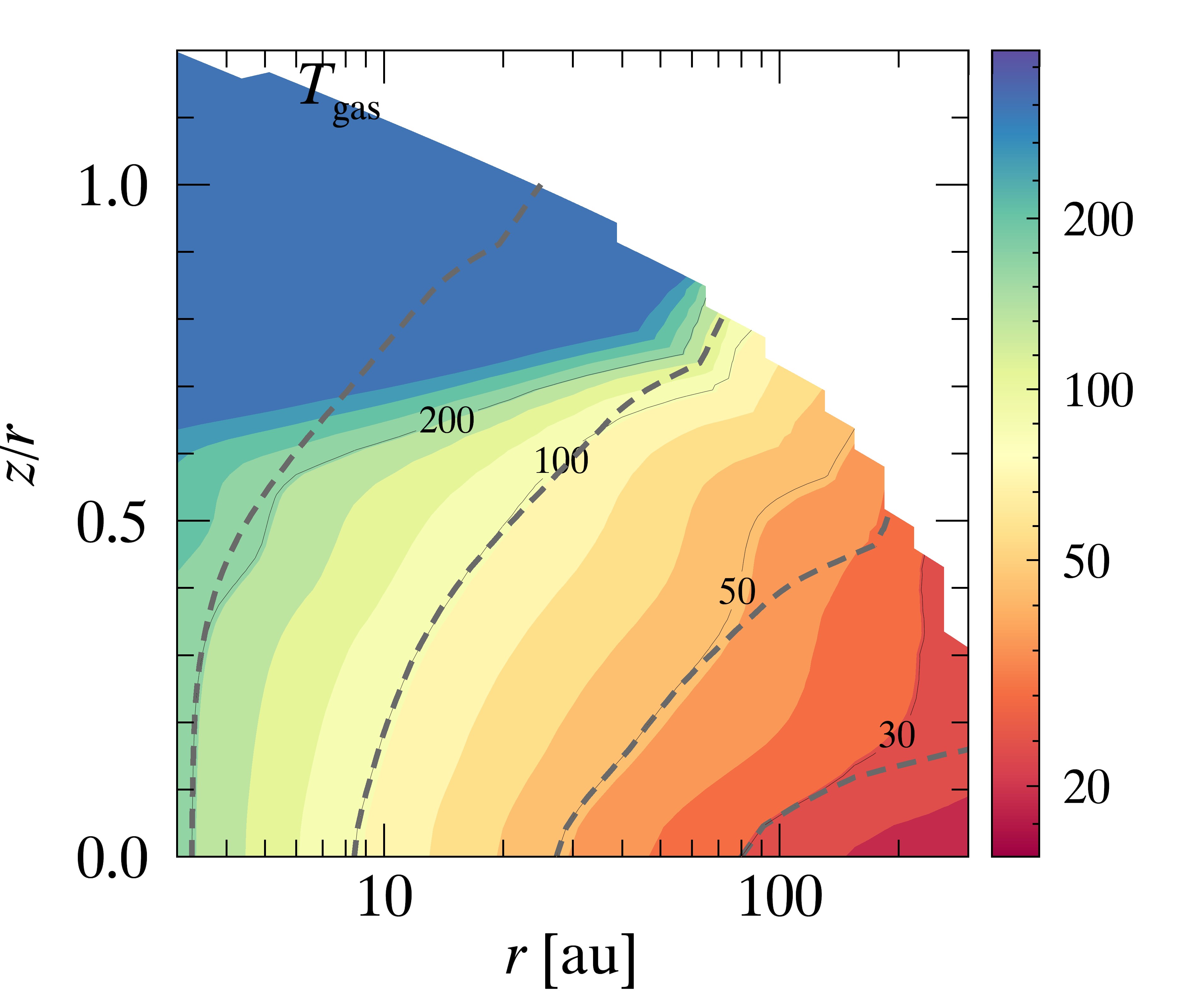}
\includegraphics[width=8cm]{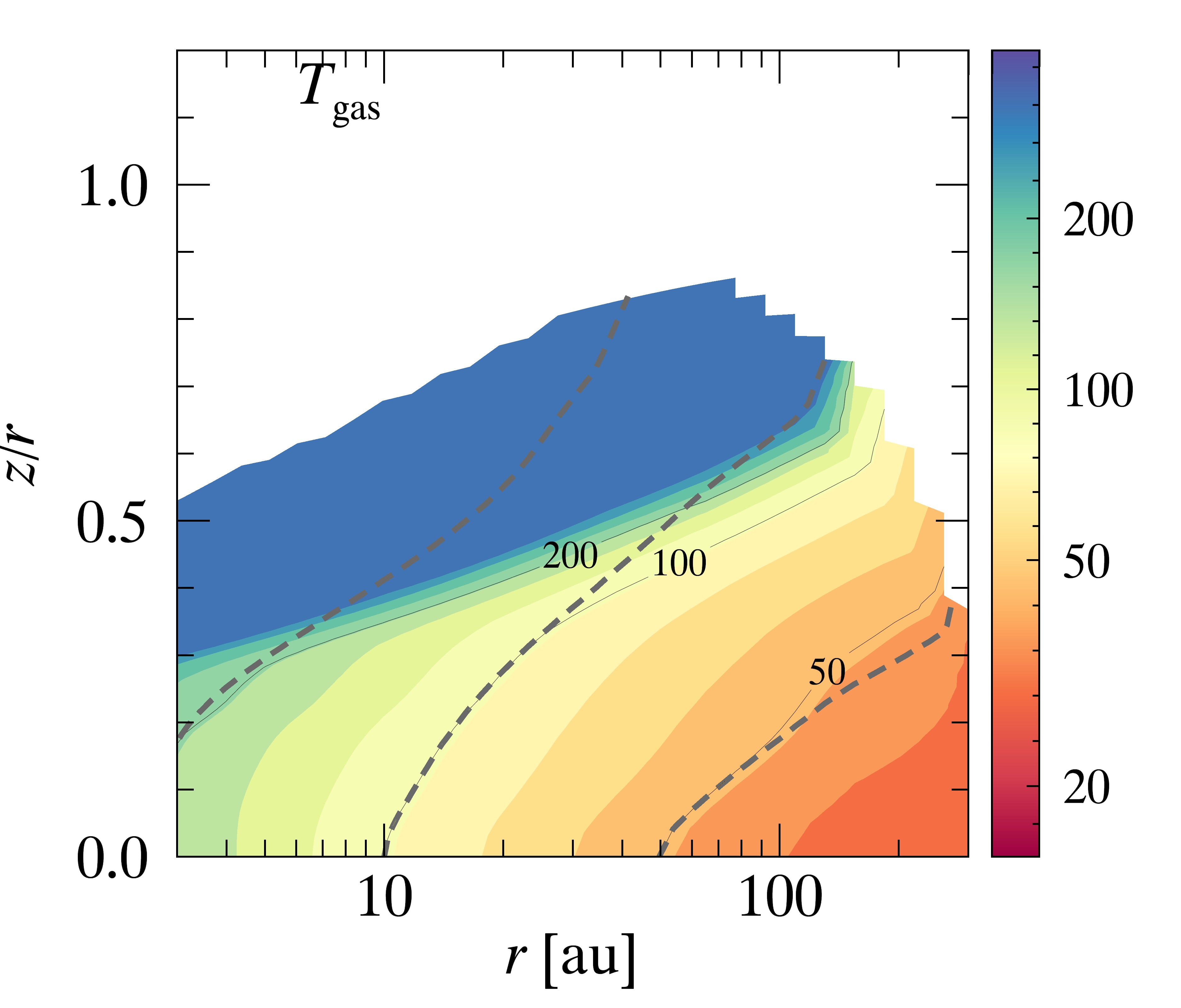}
\includegraphics[width=9cm]{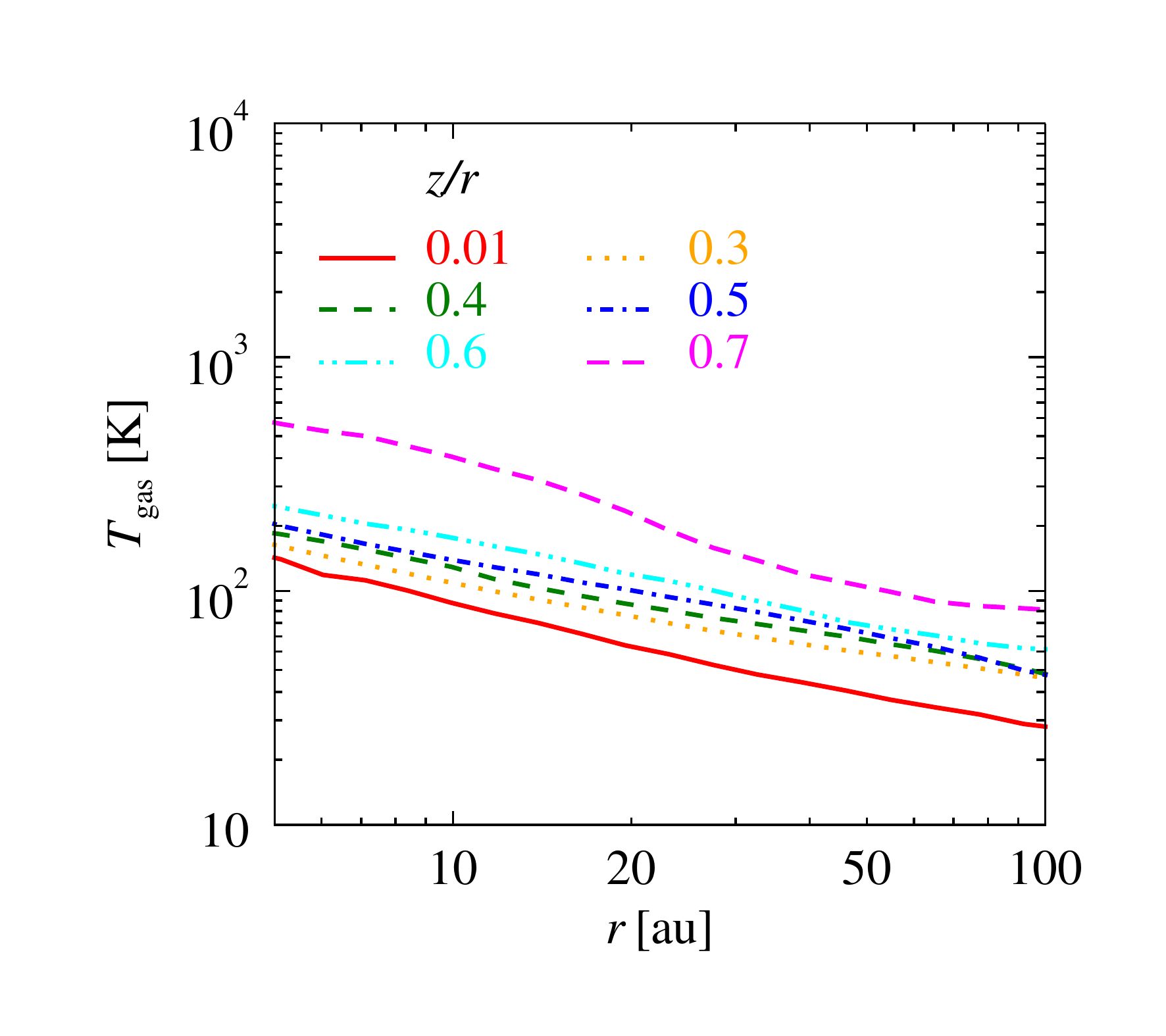}
\includegraphics[width=9cm]{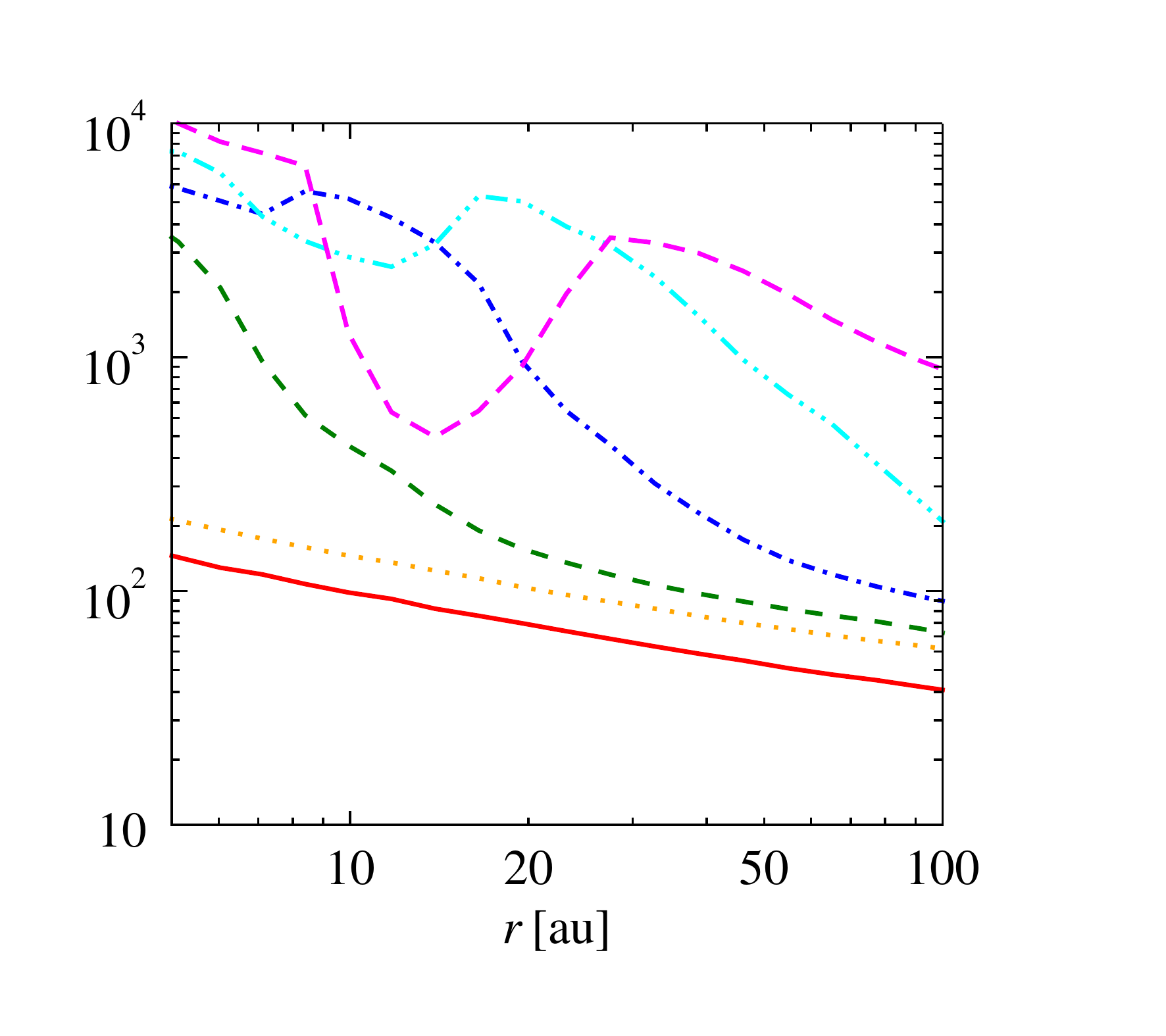}
\caption{
DALI disk structure of the two representative models, $\psi = 0.05$ (left panels) and $\psi = 0.25$ (right panels).
(Top) Gas density structure: the inset shows the inner disk structure, the dark-blue curves indicate the $n_{\rm gas}=10^6, 10^8$ and $10^{10}$\,cm$^{-3}$ contours.
The line contribution functions of a mix of low- and high$-J$ transitions of $^{12}$CO 
(solid lines) and $^{13}$CO (dashed) are overlaid on the $n_{\rm gas}$ panels. Each contour shows the layer where 50\% of the line flux emerges. 
(Middle) Gas temperature structure, the isothermal contours are overlaid for $T_{\rm gas}$ and $T_{\rm dust}$ (dashed lines) = 20, 50, 100, 200 and 500\,K.
(Bottom) Gas temperature radial profile at different disk height relevant for the disk midplane ($z/r=0.01$) and for the CO emitting layers.
}  
\label{fig:dali}
\end{figure*}

\smallskip
\noindent
Figure~\ref{fig:dali} shows the gas density (top row) structure for the two representative models (Table~\ref{tab:dali}, the spectral energy distribution of the two 
models is shown in Appendix). The inset shows the inner disk structure.
The line contribution functions of a set of low-$J$ and high-$J$ transitions of $^{12}$CO and $^{13}$CO are overlaid on the $n_{\rm gas}$ contours showing the layer where 50\% of the 
line flux emerges. The CO emitting layer varies slightly with $J$ and the four transitions shown here emerge from a layer between $z/r \sim 0.4-0.6$ and $z/r \sim 0.3-0.5$ for
the flat and flared disk, respectively. In all cases, the $^{13}$CO lines emerge closer to the disk midplane and at smaller stellar distances compared to the $^{12}$CO lines; 
because 
of the vertical and radial density gradients, the CO emission becomes optically thick higher up in the disk atmosphere and at larger distance from the star compared 
to the $^{13}$CO line. This is in excellent agreement with the broader velocity profile observed for $^{13}$CO (Table~\ref{tab:log}, Figure~\ref{fig:spectra}, bottom row).

\noindent
\smallskip
The gas temperature structure for the representative models in shown in Figure~\ref{fig:dali} (middle panel). At any given position in the disk, the gas and dust 
temperatures increase with flaring angle. Note in particular that the dust temperature in both disks is $T_{\rm dust} > 20\,$K everywhere in the disk. This prevents CO 
from freeze-out on dust grains as condensation occurs only at $T \lesssim 20\,$K in the disk interior. Without CO freeze-out, formation of complex species via surface 
chemistry is inhibited in such a warm disk. In both cases, $T_{\rm gas}$ is larger than $T_{\rm dust}$ in the upper layers of the disk. 

\begin{figure*}
\centering
\includegraphics[width=14cm]{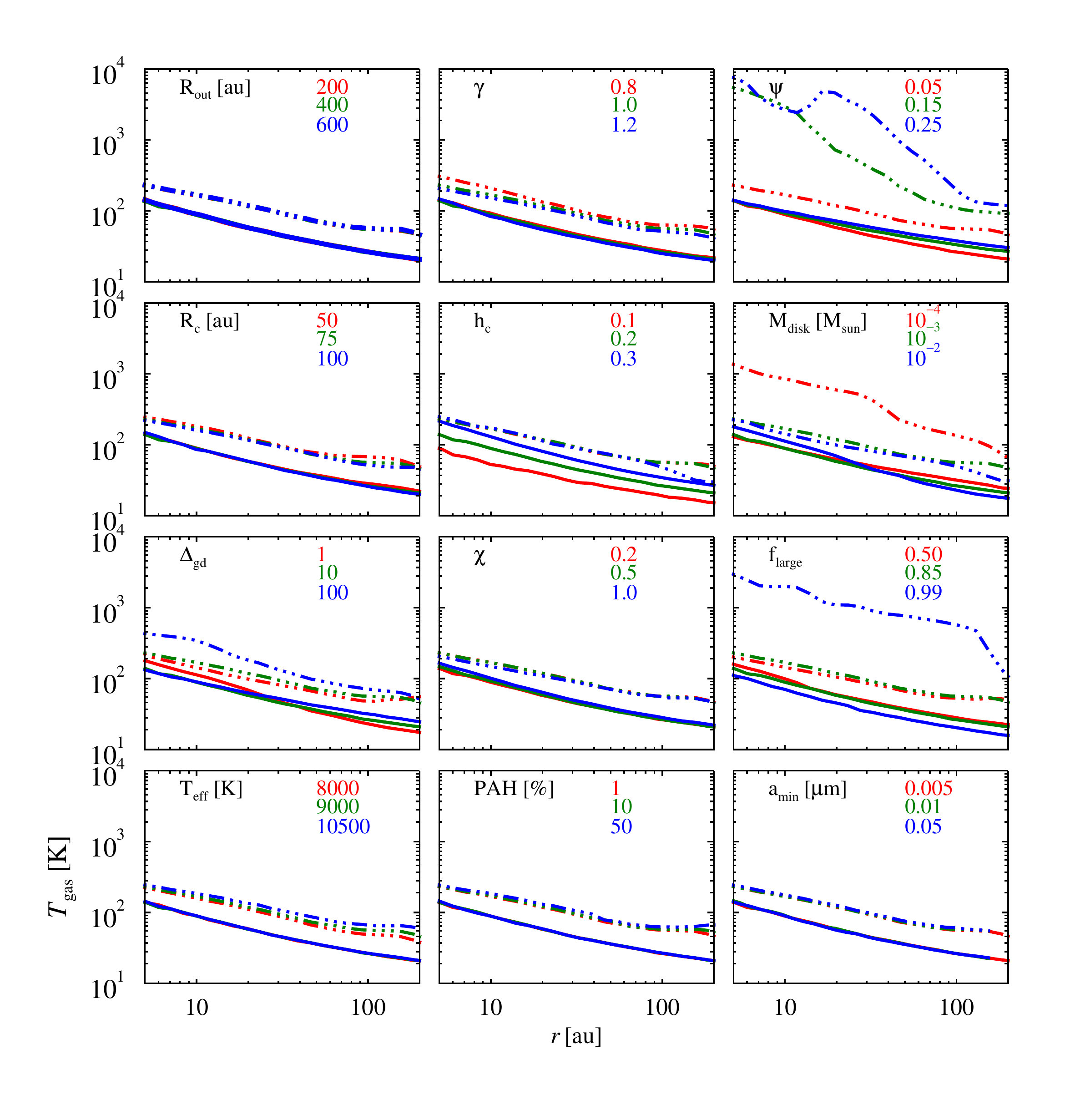}
\caption{
$T_{\rm gas}$ radial profile for different disk/stellar parameters. The temperature radial profile is shown for the layers $z/r$=0.01 (solid lines) and 0.6 (dot-dashed lines).
In the case of $h_{\rm c} = 1$, the CO emitting layer is $z/r \sim 0.2-0.3$ and the dost-dashed line shows the temperature gradient at $z/r=0.3$.}
\label{fig:tprofile}
\end{figure*}

\begin{figure*}
\centering
\includegraphics[width=14cm]{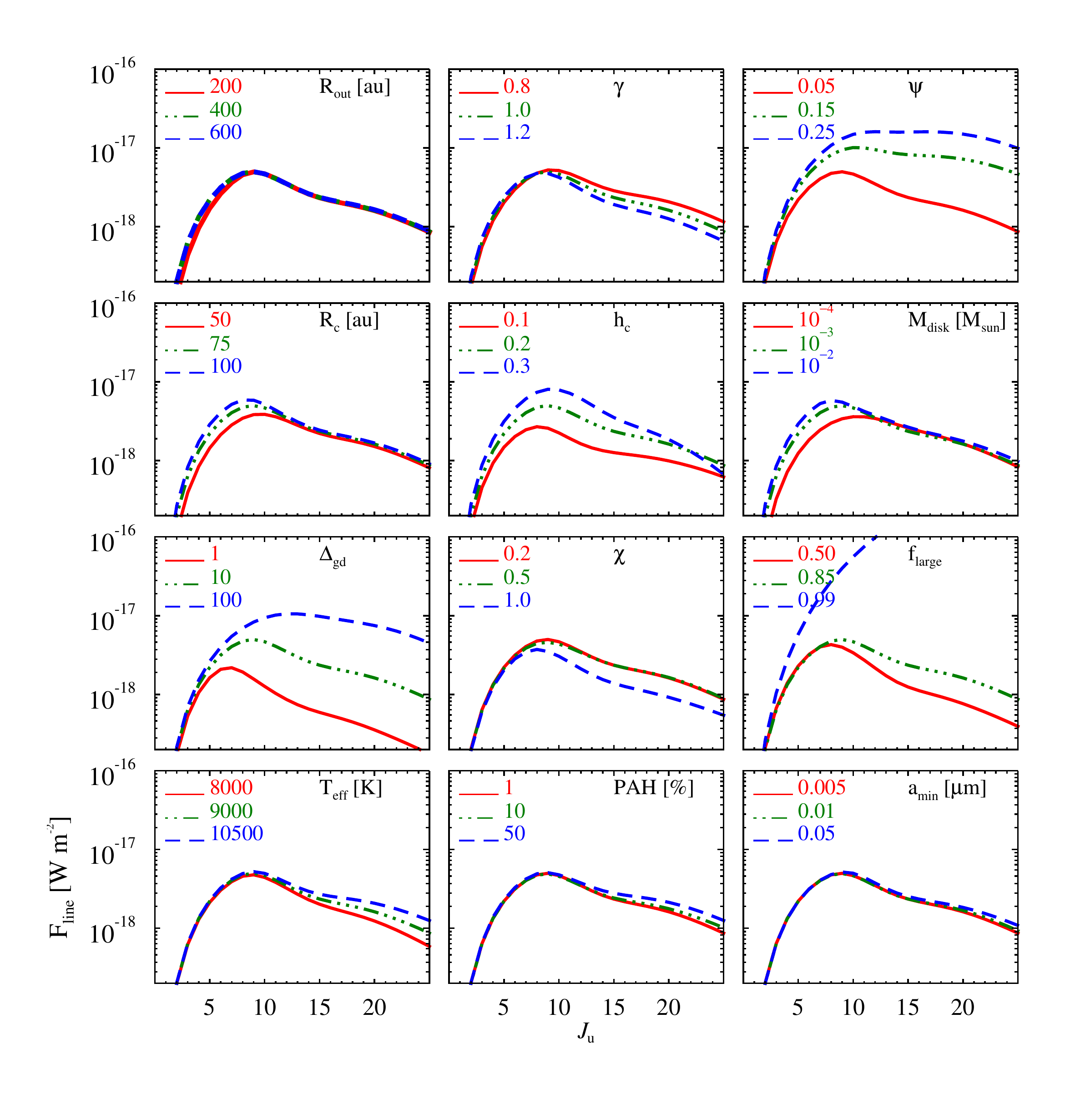}
\caption{
Synthetic CO ladder for different disk parameters (see Table~\ref{tab:dali}). Fluxes measured assuming a distance of 100\,pc and a disk inclination of 45$^{\circ}$}
\label{fig:dali_ladder}
\end{figure*}

\smallskip
\noindent
The $T_{\rm gas}$ radial profile at different height ($z/r$) for the two representative models is shown in the bottom panel of Figure~\ref{fig:dali}: in the flat disk case, the 
temperature increases slightly with height with an almost identical radial dependence from the disk midplane ($z/r=0.01$) up to the disk upper layers. This is no longer true for
flared disks which show a strong dependence of $T_{\rm gas} (r)$ with disk height. 

\subsection{Disk models grid}
This section provides a qualitative comparison to physical-chemical models of Herbig AeBe disks. 
For this, a grid of DALI disk models is built to study the impact of different stellar and disk parameters on the disk thermal structure.
The stellar mass and the bolometric luminosity are taken to be fixed while the stellar effective temperature ($T_{\rm eff}$, assuming black body emission), 
the disk mass ($M_{\rm disk}$), critical radius ($R_{\rm c}$), disk outer radius ($R_{\rm out}$), scale height ($h_{\rm c}$), power-law exponent ($\gamma$), 
flaring angle ($\psi$), dust settling ($\chi, f_{\rm large}$), minimum grain size ($a_{\rm min}$), gas-to-dust mass ratio ($\Delta_{\rm gd}$) and the PAH 
abundances are varied (Table~\ref{tab:dali}). Several of these parameters were also investigated in \citet{Bruderer12} but only for a flared disk. 
The model grid is built around the flat ($\psi = 0.05$) disk representative model varying the aforementioned parameters by the values given in (Table~\ref{tab:dali}). 
In particular, the values of $T_{\rm eff}$ are representative of the typical stellar temperature of Herbig Ae systems \citep[e.g.,][]{vandenancker98}.
The FUV ($6-13.6\,$eV) luminosity, relevant for the heating of the disk and the photodissociation of CO, is regulated by $T_{\rm eff}$ and it ranges between
$L_{\rm FUV}=0.45 - 1.7\,L_{\odot}$ for $T_{\rm eff}=8,000 - 10,500$\,K, respectively.

\smallskip
\noindent
To investigate the impact of dust settling on the disk thermal structure, $\chi$ and $f_{\rm large}$ are varied between $0.2-1,0$ and $0.5-0.999$, 
respectively (Table~\ref{tab:dali}). Lowering the value of $\chi$ has the effect to let the UV photons penetrate further inside the disk. As a 
consequence the C$^+$/C/CO transition layers shift deeper into the disk. For $f_{\rm large}$ = 0.5 the dust mass is distributed equally between the small
and the large dust grains, for $f_{\rm large}$ = 0.999, the bulk of the dust mass is in the large dust grains.

\smallskip
\noindent
The abundance of PAHs and the gas-to-dust mass ratio are poorly constrained in disks. Model fits to observational data suggest PAH abundances that
are typically 10-100 times lower than those in the interstellar medium \citep{Geers06}. For Herbig AeBe disks, PAHs of about 100 carbon atoms survive 
the strong UV radiation \citep{Visser07}. For this analysis three abundance values are considered, 1\% 10\% and 50\% with respect to the abundance 
of the interstellar medium \citep{Draine07}. The gas-to-dust mass ratio ($\Delta_{\rm gd}$) in disks is likely to be lower than that of the interstellar 
medium \citep[e.g.][]{Chapillon10}. In this paper $\Delta_{\rm gd}$ varies between 1\% and 100\% the ISM value \citep{Bohlin78}.

\noindent
The effects of elemental carbon abundance and isotope selective photodissociation (relevant for $^{13}$CO) are not treated here. These are investigated in \citet{Bruderer12} and \citet{Miotello14}. 
As noted by \citet{Miotello14}, the $^{13}$CO line intensities are less affected by the isotope selective photodissociation compared to C$^{18}$O and C$^{17}$O. It is thus reasonable to neglect this 
effect in this analysis. The effect of a lower carbon abundance is similar to that of a smaller gas-to-dust ratio \citep{Bruderer12}.

\subsection{Temperature radial gradient}
Figure~\ref{fig:tprofile} shows the $T_{\rm gas}$ radial profile for all disk models: each panel shows the temperature gradient in the disk midplane ($z/r = 0.01$) and at $z/r=0.6$,  
representative of the CO emitting layer. In the case of $h_{\rm c} = 1$, the CO emitting layer is $z/r \sim 0.2-0.3$ and the dost-dashed line shown in Figure~\ref{fig:tprofile} shows the 
temperature gradient at $z/r=0.3$.

\smallskip
\noindent
The parameters affecting most the midplane temperature are the flaring angle, scale height, the dust settling and to a lesser extent the disk mass and the gas-to-dust
mass ratio. At the CO emitting layer the temperature gradient is controlled mostly by the disk flaring and to a less extent by the disk mass, the gas-to-dust mass ratio, the 
dust settling ($f_{\rm large}$) and by the stellar temperature.    
Note that the $f_{\rm large}=0.999$ case is a very extreme scenario in which the disk outer layers are almost completely devoid of dust as the bulk of the mass is the
large dust grains, which are settled in the disk midplane ($\chi=0.2$).

\subsection{CO line fluxes}
The synthetic CO ladders are shown in Figures~\ref{fig:dali_ladder}. In all cases the distance is fixed to 100\,pc and the disk inclination to 45$^{\circ}$. 
%The aim of this paper is not to model individual sources but rather explain the trends observed in Figure~\ref{fig:coladder}.

\smallskip
\noindent
There are two main aspects of the CO ladder that the models need to reproduce: the absolute line fluxes and the shape of the CO ladder, in particular whether 
or not it bends over at higher $J$. No changes are observed for different disk size. The parameter that affects primarily the CO line ladder are:

\begin{description}
\item[Flaring angle]: the CO line fluxes vary by 1-2 order of magnitude between the flat (fainter) and the flared disks for $J \gtrsim 10$ and the line flux difference 
increases with $J$. This is due to the different gas temperature structure between flat and flared disks (Figure~\ref{fig:tprofile}) \\

\item[Critical radius]: the low$-J$ ($J < 10$) lines become brighter with $R_{\rm c}$ while the flux of the high$-J$ does not vary. Since the temperature structure does 
not vary with $R_{\rm c}$ (Figure~\ref{fig:tprofile}), this is likely due to the change in vertical depth in the outer disk due to the change in $R_{\rm c}$ \\

\item[Scale height]: all CO lines become brighter if the scale height increases. Since the temperature radial gradient does not show significant changes with $h_{\rm c}$ 
(at the CO emitting layer), the differences in CO line fluxes are due to shift of the C$^+$/C/CO transition layers deeper into the disk with decreasing scale height \\

\item[Disk mass]: it affects the low$-J$ ($J < 10$) lines only which become slightly brighter for increasing disk mass. The effect is only significant for low disk masses 
($\lesssim 10^{-4}\,M_{\odot}$) and it is due to the change in gas column density (hence line optical depth) in the outer disk \\

\item[Gas-to-dust mass ratio]: reducing $\Delta_{\rm gd}$ from 100 to 1 has the effect of lowering all line fluxes by up to one order of magnitude. Note that keeping the 
gas mass constant and lowering the gas-to-dust mass ratio implies a higher dust mass. Thus the CO lines are fainter for low values of  $\Delta_{\rm gd}$ because of the 
increased opacity and lower temperature (Figure~\ref{fig:tprofile}) induced by the higher dust mass compared to the $\Delta_{\rm gd}=100$ case \\

\item[Dust settling and large-to-small grains mass ratio]: the high$-J$ lines are brighter for a settled disk and for an higher mass ratio. This is due to the increase
in gas temperature in the disk upper layers. In the extreme case of $f_{\rm large} = 0.999$ all the CO lines are several order of magnitudes brighter. 

\end{description}

\smallskip
\noindent
All other remaining parameters analysed here affect mostly the flux of the high$-J$ ($J \gtrsim 10$) lines inducing a bending over the CO ladder for $J \gtrsim 10$. 

\begin{figure}
\centering
\includegraphics[width=7cm]{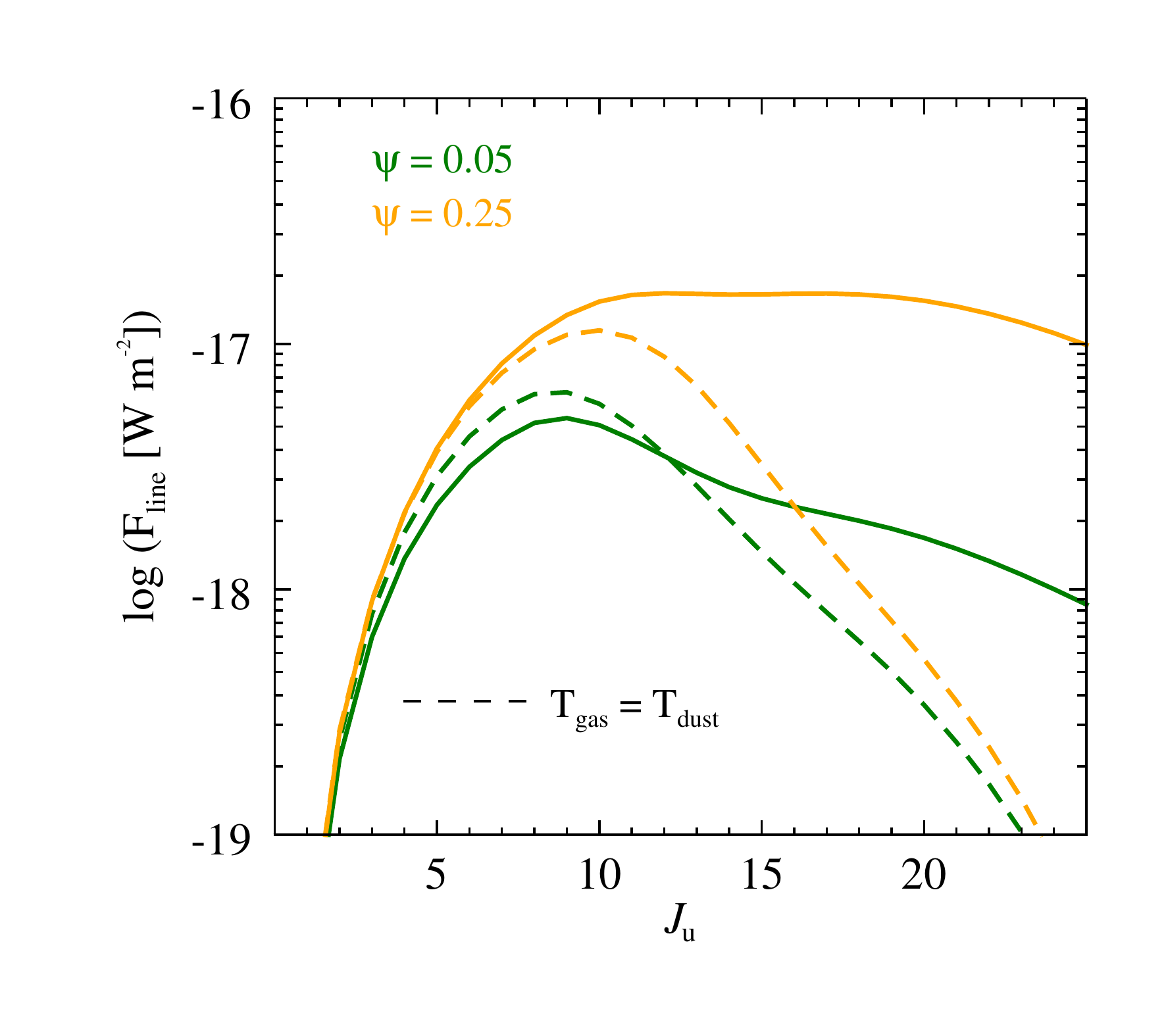}
\caption{Synthetic CO ladder in the case of $T_{\rm gas} = T_{\rm dust}$ case (representative models only).}
\label{fig:notherm}
\end{figure}

\smallskip
\noindent
For comparison, Figure~\ref{fig:notherm} shows the CO ladder in the case of a ``cold" disk atmosphere (i.e. $T_{\rm gas} = T_{\rm dust}$, obtained by switching 
off the thermal balance in \textsc{DALI}). In this case all the CO line fluxes are substantially lower and the flux drops quickly with $J$ for $J_{\rm u} \gtrsim 10$. 
The shape of the observed CO ladder and the absolute flux level are inconsistent with the cold disk case, thus suggesting that $T_{\rm gas}$ exceeds $T_{\rm dust}$ in 
the upper layers of all the Herbig Ae disks studied here.

\begin{figure}
\centering
\includegraphics[width=7cm]{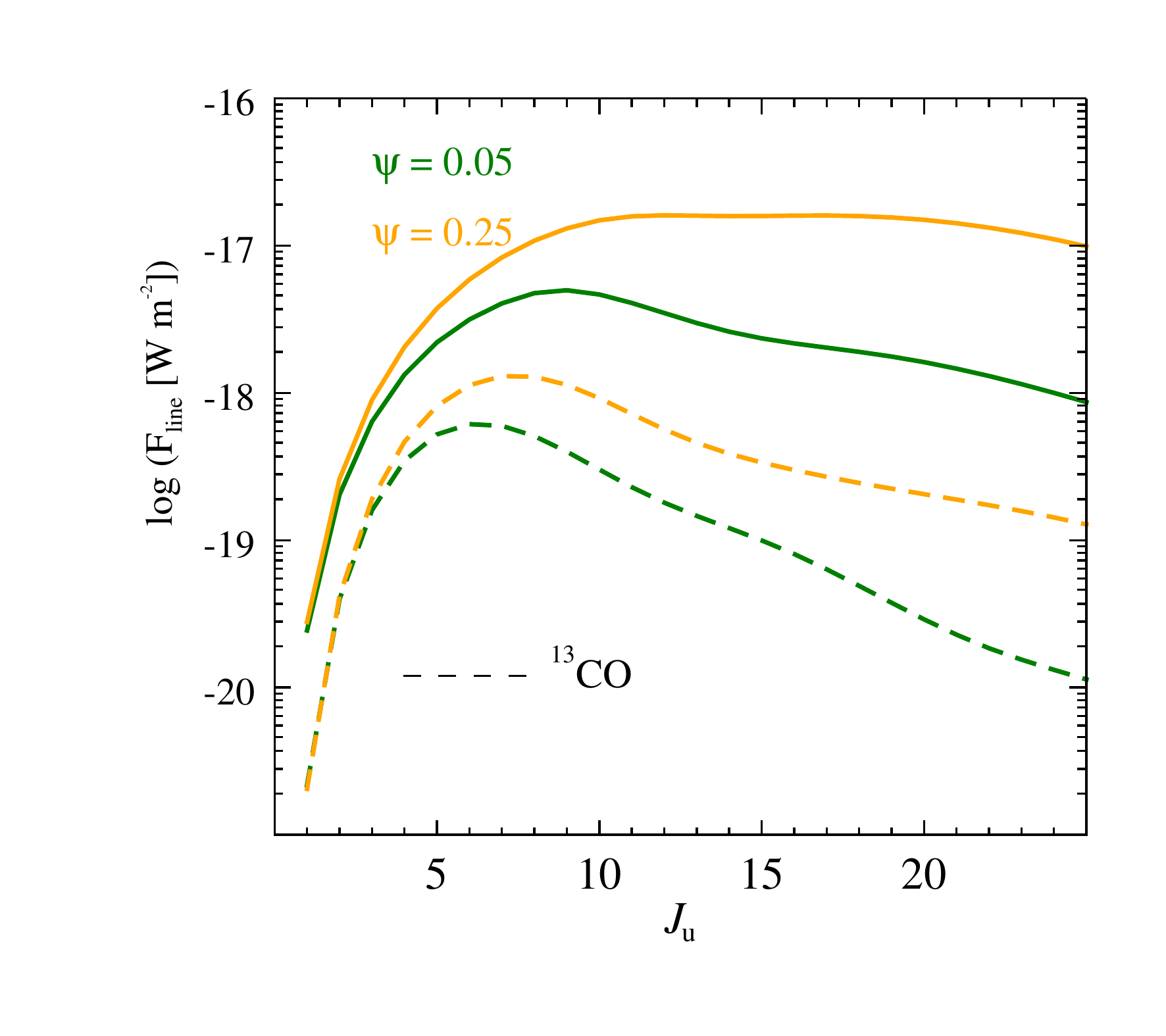}
\caption{Synthetic $^{13}$CO ladder (representative models only).}
\label{fig:13co}
\end{figure}

\smallskip
\noindent
The $^{13}$CO ladder is shown in Figure~\ref{fig:13co} for two representative disk models. The fluxes of the $^{13}$CO lines in the PACS
range are all below the detection threshold of the DIGIT and GASPS surveys. The estimate of the $^{13}$CO $J=10-9$ flux is in good 
agreement (once corrected for distance) with the value measured in this paper for HD 100546 and HD 163296 (Table~\ref{tab:log}). The fluxes of the mid-$J$ $^{13}$CO 
lines reported by \citet{vanderwiel14} toward HD 100546 (based on Herschel/SPIRE) are much brighter than the values estimated here. As
noted by \citet{vanderwiel14} the SPIRE measurements may be contaminated by an extended, non-disk, emission.

\smallskip
\noindent
As shown by the line contribution function in Figure~\ref{fig:dali}, 
the $^{13}$CO emission probes a layer closer to the disk midplane at smaller distance to the star compared to the $^{12}$CO emission. The relative 
fluxes and line profiles of multiple high-$J$ transitions of the two isotopologues provide a strong constraint to the disk temperature structure 
both in the radial and in the vertical direction.

\begin{figure}
\centering
\includegraphics[width=7cm]{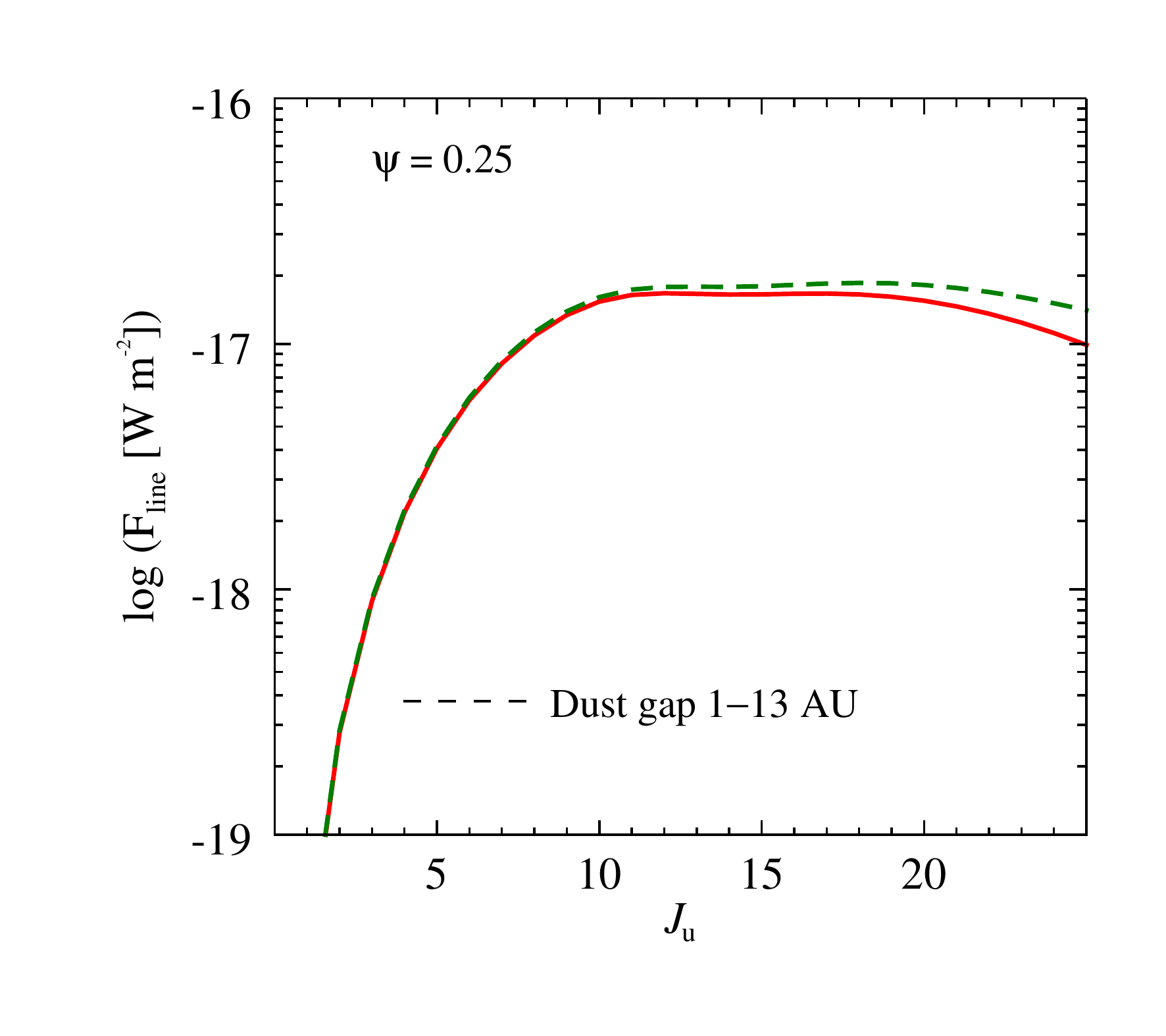}
\caption{Synthetic CO ladder for the flared disk case ($\psi=0.25$) with (dashed, green curve) and without dust gap.}
\label{fig:gap}
\end{figure}

\smallskip
\noindent
Finally, Figure~\ref{fig:gap} shows the effect of a dust gap in the CO ladder for the flared ($\psi=0.25$) disk case: 
a dust gap is included between $1-13$\,au with a drop in dust surface density of $\delta_{\rm dust} = 10^{-6}$ and a drop in the gas surface density $\delta_{\rm gas} = 10^{-6}$.
Such a structure is representative of the Herbig Ae group I systems analyzed here (e.g., HD 100546). As the figure shows, the dust gap does not impact the CO ladder for $J < 25$. This
is because the high$-J$ lines analysed here are emitted at radii larger than the dust gap.

\subsection{CO line profiles}
\begin{figure*}
\centering
\includegraphics[width=16cm]{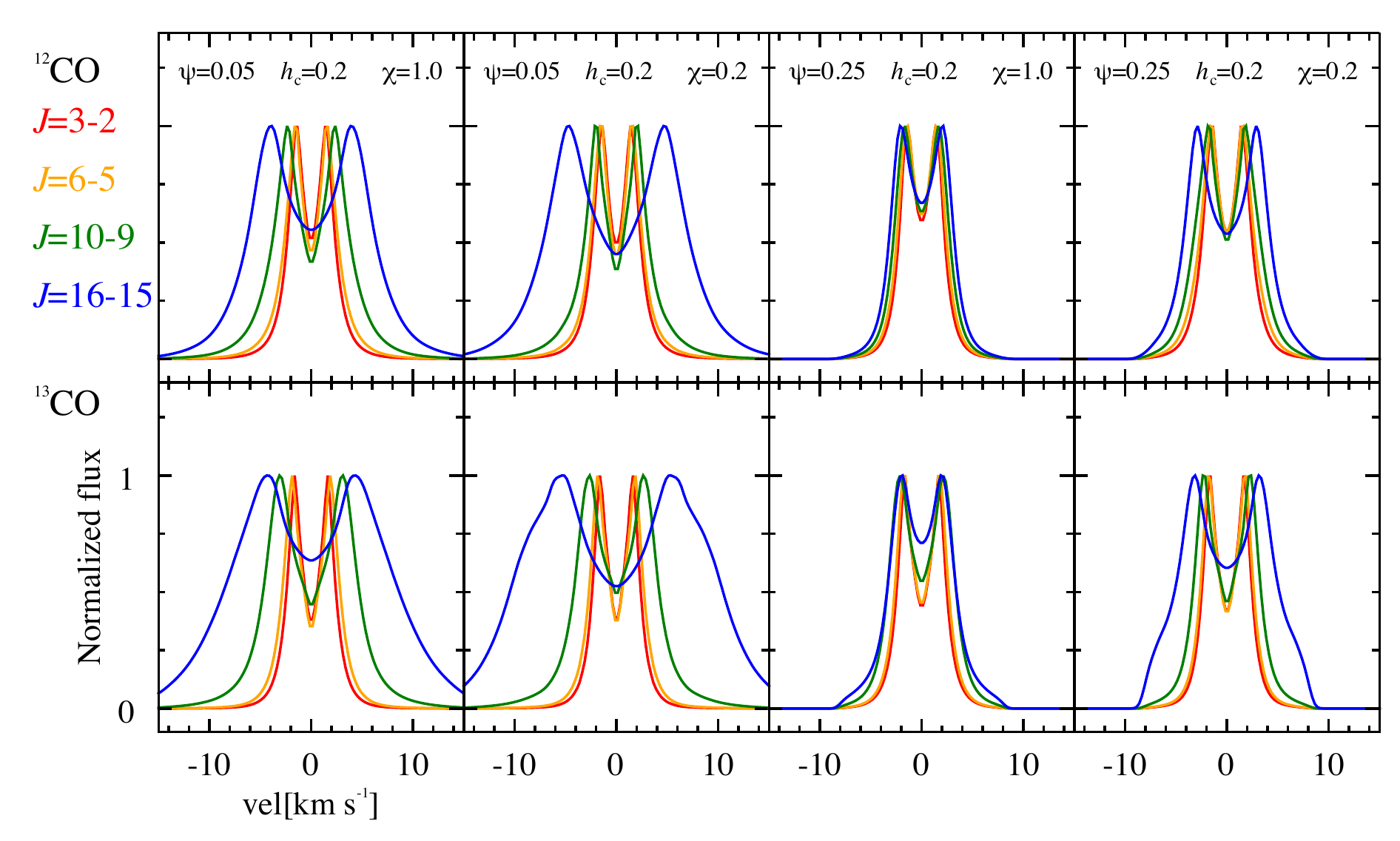}
\caption{CO (top) and $^{13}$CO (bottom) line profiles computed with {\textsc DALI} as a function of dust settling and flaring angle for a disk inclination of $i=45^{\circ}$. The remaining parameters are fixed:
h$_{\rm c}$=0.2, $\Delta_{\rm gd}=100$, $T_{\rm eff} = 10,500\,$K,  PAH=1\%, $f_{\rm large}=0.85$.}
\label{fig:dali_profile}
\end{figure*}

The predicted velocity profiles of different CO transitions are shown in Figure~\ref{fig:dali_profile} ($^{13}$CO) for the representative models. 
A mix of low- and high-$J$ transitions are plotted for a disk inclination of $i = 45^{\circ}$. In all cases the line width increases with $J$. The larger differences in line width 
are seen for the flat disk models. 
Note that these differences in line width become particularly significant for $J=16-15$, not yet for $J=10-9$, demonstrating the power of these high-$J$ HIFI data.
For $\psi=0.05$ the disk scale height and the dust settling have only minor effect on the line velocity profiles. In the case of 
$\psi=0.25$ instead, all the lines become broader if there is substantial dust settling while changing the scale height from $h_{\rm c}=0.2$ to 0.3 has little impact (not shown here). 
 
\smallskip
\noindent
The $^{13}$CO lines are systematically broader than the $^{12}$CO lines (Figure~\ref{fig:dali_profile}) and this is particularly true for the higher $J$ transitions.
This is because the $^{13}$CO transitions become optically thick at higher density and the line emitting region moves deeper into the disk and closer to the star 
(for a given $J$) compared to the $^{12}$CO transitions. This can be seen in Figure~\ref{fig:dali} which shows the gas density structure and the line contribution function.

\section{Discussion and conclusion}
The major finding of this paper is that the disk temperature structure varies substantially from system to system (Figure~\ref{fig:tgas}). This result is based on 
the analysis of the high-$J$ CO line profiles and of the CO rotational ladder.
Note in particular that the 4 Herbig Ae systems have similar stellar mass ($M_{\odot} \sim 2-2.5\,M_{\odot}$) while the disk 
temperature can differ by almost an order of magnitude.
The comparison of the observed (Figure~\ref{fig:coladder}) and synthetic (Figure~\ref{fig:dali_ladder}) CO ladders suggest a disk temperature sequence for the
4 Herbig Ae systems studied here (when accounted for the different distances of the sources). 
In this sequence  HD 100546 is the hottest and HD 163296 the coldest disk. The primary parameters regulating the overall disk temperature are the flaring angle, 
the scale height and the the gas-to-dust mass ratio; varying these parameters induce variation of the CO line flux up to 1-2 order of magnitude. 
The disk temperature structure is sensitive to the disk mass only for $M_{\rm disk} \lesssim 10^{-4}\,M_{\odot}$. Dust settling may also lead to changes in the disk temperature. 
The power-law exponent of the surface density profile, the stellar temperature and PAH abundance have secondary effects on the temperature of Herbig Ae disks 
affecting mostly the CO ladder shape, producing a bending in the CO rotational ladder for $J \gtrsim 10$.
In some cases, changes in the CO ladder are not due to a different temperature structure (Sec.~5.3): as an example, a different disk critical radius affects the 
value of the low-$J$ CO lines ($J < 10$) while the overall temperature structure does not vary substantially.

\smallskip
\noindent
An interesting outcome of the analysis performed here is that the CO ladder may help to constrain some key physical parameters in disks like
the gas-to-dust mass ratio, or equivalently the elemental carbon abundance \citep{Bruderer12}. Spatially-resolved millimetre images of the low-$J$ lines 
allow to accurately measure the disk flaring angle and scale height \citep{degregorio13} while multi-frequency dust continuum observations 
provide strong constraints on the grain size distribution reducing the degeneracy in the CO ladder. Once these properties are measured, 
the major differences in the CO ladder are driven by the gas-to-dust mass ratio.   

\smallskip
\noindent
In conclusion, the CO rotational ladder and the velocity profiles of multiple-$J$ transitions are a valid diagnostic of $T_{\rm gas}$ in disks. 
With the end of 
{\it Herschel} operation, the investigation of the far-infrared spectrum of disks is currently limited to the very bright sources observable with SOFIA. Future 
observations with, e.g., cryogenic facilities like SPICA will allow to expand the analysis of the CO (and isotopologues) rotational ladder to a much larger sample of
protoplanetary systems allowing to routinely measure the temperature structure of not only flared but also flat Herbig AeBe and T Tauri disks. In particular, the line flux ratio of several, 
mid- and high$-J$ CO and $^{13}$CO will provide direct insights on the vertical (different $\tau$) and radial (different $J$) temperature structure of protoplanetary
disks if complemented by lower-$J$ spectrally and spatially resolved data from, e.g., ALMA. 

\begin{acknowledgements}
DF acknowledges support from the Italian Ministry of Science and Education (MIUR), project SIR (RBSI14ZRHR).
The authors thanks the WISH team and L. Podio for providing the HIFI $^{12}$CO and $^{13}$CO $J=10-9$ spectra of TW Hya and HD 163296. 
We are also grateful to the DIGIT, GASPS teams for providing the PACS spectra of CO.
Astrochemistry in Leiden is supported by the Netherlands Research School for Astronomy (NOVA), by a Royal Netherlands Academy of Arts and Sciences (KNAW) professor
prize, and by the European Union A-ERC grant 291141 CHEMPLAN.
\end{acknowledgements}

%\bibliographystyle{aa}
%
%\bibliography{mybib}

\begin{appendix}

\section{CO Rotational Ladder}
The fluxes of the CO rotational lines up to $J=25-24$ are listed in Table~\ref{tab:coladder}. Data are from this
work and from \citet{Bergin13, Bruderer14, Fedele13b, Kamp13, Meeus13, Qi11, vanderwiel14} (and references therein) 
and from Kama et al. (submitted). The flux of the $J=6-5$ in IRS 48 is a lower limit because the line is partially 
obscured by the molecular cloud \citep{Bruderer14}. Some of the high-$J$ line fluxes ($J > 14-13$) based on 
{\it Herschel}/PACS observations have been re-measured with the most updated instrument's calibration and some of 
the fluxes are slightly different than previous measurements. The $J=24-23$ in TW Hya is blended with H$_2$O $4_{14}-3_{03}$.

\begin{table*}
\centering
\caption{CO rotational ladder fluxes}
\begin{tabular}{llllll}
\hline\hline
      & TW Hya             & HD 97048      & HD 100546    & IRS 48 & HD 163296 \\
      & \multicolumn{5}{c}{Integrated Flux [10$^{-17}$\,W\,m$^{-2}$]}     \\
\hline
 2-1  & 0.014 $\pm$ 0.0007 &               &               &                 & 0.035 $\pm$ 0.00039   \\
 3-2  & 0.043 $\pm$ 0.0013 &               & 0.15$\pm$0.01 &                 & 0.125 $\pm$ 0.00130   \\
 4-3  &                    &               &               &                 &                       \\
 5-4  &                    &               &               &                 & 1.04  $\pm$ 0.40      \\
 6-5  & 0.13 $\pm$ 0.03    & 0.63$\pm$0.03 & 1.3 $\pm$ 0.1 & $> 0.17$        & 0.74  $\pm$ 0.29      \\
 7-6  &                    & 0.8 $\pm$ 0.4 &               &                 & 0.90  $\pm$ 0.30      \\
 8-7  &                    & 1.8 $\pm$ 0.8 &               &                 & 1.24  $\pm$ 0.55      \\
 9-8  &                    & 2.4 $\pm$ 0.5 &               &                 & 0.91  $\pm$ 0.40      \\
10-9  & 0.28 $\pm$ 0.06    & 3.5 $\pm$ 0.7 & 4.5 $\pm$ 0.1 &                 & 1.17  $\pm$ 0.35      \\
11-10 &                    & 3.6 $\pm$ 0.7 &               &                 & 1.13  $\pm$ 0.35      \\
12-11 &                    & 2.8 $\pm$ 0.6 &               &                 & 1.17  $\pm$ 0.35      \\
13-12 &                    & 4.9 $\pm$ 0.9 &               &                 & 1.52  $\pm$ 0.40      \\
14-13 &                    & 2.4 $\pm$ 1.4 & 8.5 $\pm$ 1.0 & $<$ 1.8         & $<$ 1.60              \\
15-14 &                    & 3.0 $\pm$ 0.8 & 6.9 $\pm$ 1.0 & 1.3 $\pm$ 0.5   & 1.03  $\pm$ 0.50      \\
16-15 &                    & 3.0 $\pm$ 0.5 & 5.6 $\pm$ 1.0 & 1.6 $\pm$ 0.4   & $<$ 1.30              \\
17-16 &                    & 2.6 $\pm$ 0.5 & 7.1 $\pm$ 1.0 & 1.7 $\pm$ 0.4   & 0.75  $\pm$ 0.25      \\
18-17 & 0.30 $\pm$ 0.10    & 2.6 $\pm$ 0.5 & 5.8 $\pm$ 1.0 & 1.3 $\pm$ 0.3   & $<$ 0.90              \\
19-18 &                    & $<$ 2.3       & 7.4 $\pm$ 1.0 & 1.1 $\pm$ 0.4   & $<$ 0.90              \\
20-19 &                    & $<$ 2.3       & 5.2 $\pm$ 1.0 & 1.2 $\pm$ 0.4   & $<$ 0.90              \\
21-20 &                    & $<$ 2.3       & 8.6 $\pm$ 1.0 & 1.0 $\pm$ 0.3   & $<$ 0.90              \\
22-21 &                    & $<$ 2.3       & 7.3 $\pm$ 1.0 & $<$ 1.4         & $<$ 0.90              \\
23-22 & $<$ 0.44           & $<$ 2.3       & 7.8 $\pm$ 1.1 & $<$ 1.3         & $<$ 0.90              \\
\hline                                 
\end{tabular}\label{tab:coladder}
\end{table*}

\section{\cii}
\begin{figure}
\centering
\includegraphics[width=6cm]{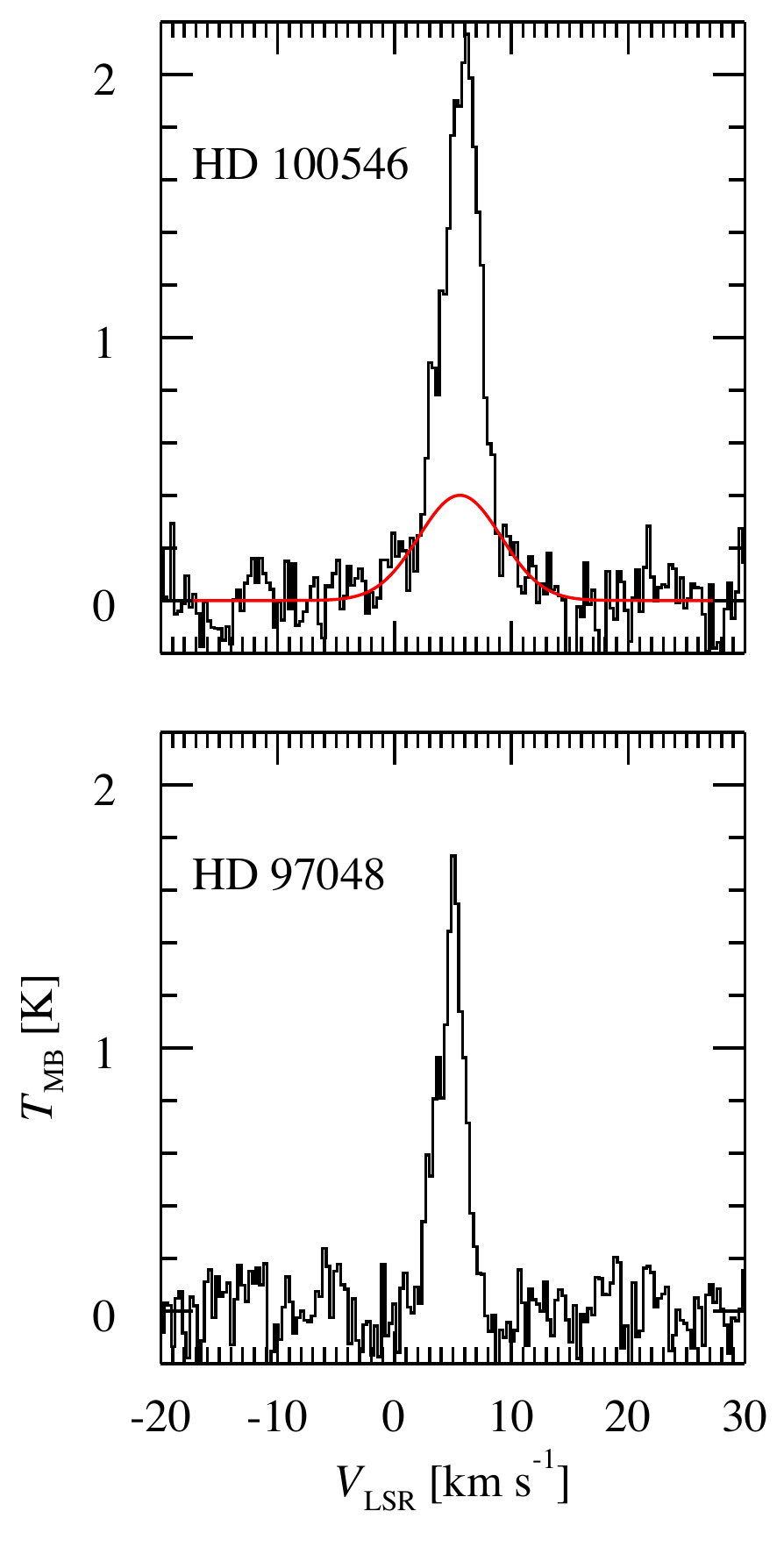}
\caption{HIFI/WBS spectra of \cii. The base of the \cii \ line in HD 100546 is fitted with a Gaussian profile (red curve). }
\label{fig:cii}
\end{figure}

\cii \ emission is detected at high S/N toward both sources (HD 97048 and HD 100546, Figure~\ref{fig:cii}). The line is narrow with 
a Gaussian FWHM of $\sim 3-4\,$\kms.  In the case of HD 97048 the peak of the emission is at v$_{\rm LSR} \sim$ 5.1\,\kms \ , red-shifted 
from the center of the CO $J=16-15$ line by $\sim 0.2\,$\kms. 

\smallskip
\noindent
The analysis of the \cii \ profile toward HD 100546 is presented in paper I and the spectrum is shown here for comparison with that of HD 97048. 
Because of the internal load chop used in combination with an off-source calibration with HIFI, some of the large scale line emission is removed 
in the HIFI spectra. Given the large beam of HIFI at this frequency (11\farcs1), the large-scale, non-disk, emission contributes most of the \cii \ 
flux and we conclude that the emission measured here is dominated by a diffuse, low-velocity gas not associated with the disk, i.e., diffuse cloud 
or remnant envelope on scales of 1000\,au.

\smallskip
\noindent
In the case of HD 100546, the base of the \cii \ line is broader than the core of the line as can be seen from the high velocity wings. This 
broadening may be due to an underlying disk contribution. To quantify this, the line wings are fitted with a Gaussian profile centered at the 
system velocity ($V_{\rm LSR} = 5.6\,$\kms). The best fit profile is shown in Figure~\ref{fig:spectra} (red curve). The fit is partly degenerate 
in the Gaussian height and width. Nevertheless, in order to fit the high velocity wings, the FWHM must range between $\sim 7-8\,$\kms. The disk \cii \ 
contribution is estimated by integrating the flux below the best-fit Gaussian profile and we find a value of $\sim 8-9\times10^{-17}\,$W\,m$^{\rm -2}$, 
correspondent to 4-5\% of the total line flux. This value must be taken with caution for various reasons (e.g., the intrinsic line profile is not Gaussian). 

\smallskip
\noindent
No evidence of line broadening is detectable in the HIFI \cii \ spectrum of HD 97048. The upper limit to the \cii \ disk contribution is
measured assuming a line peak equal to 0.3\,K ($3 \times $rms, Table~\ref{tab:log}) and a line width of 7\,\kms \ (similar to HD 100546). This gives a $3\,\sigma$
upper limit of 2.4\,K\,\kms \ equivalent to $5.5\times10^{-17}\,$W\,m$^{\rm -2}$. 

\section{Spectral energy distribution of representative models}

\begin{figure}
\centering
\includegraphics[width=9cm]{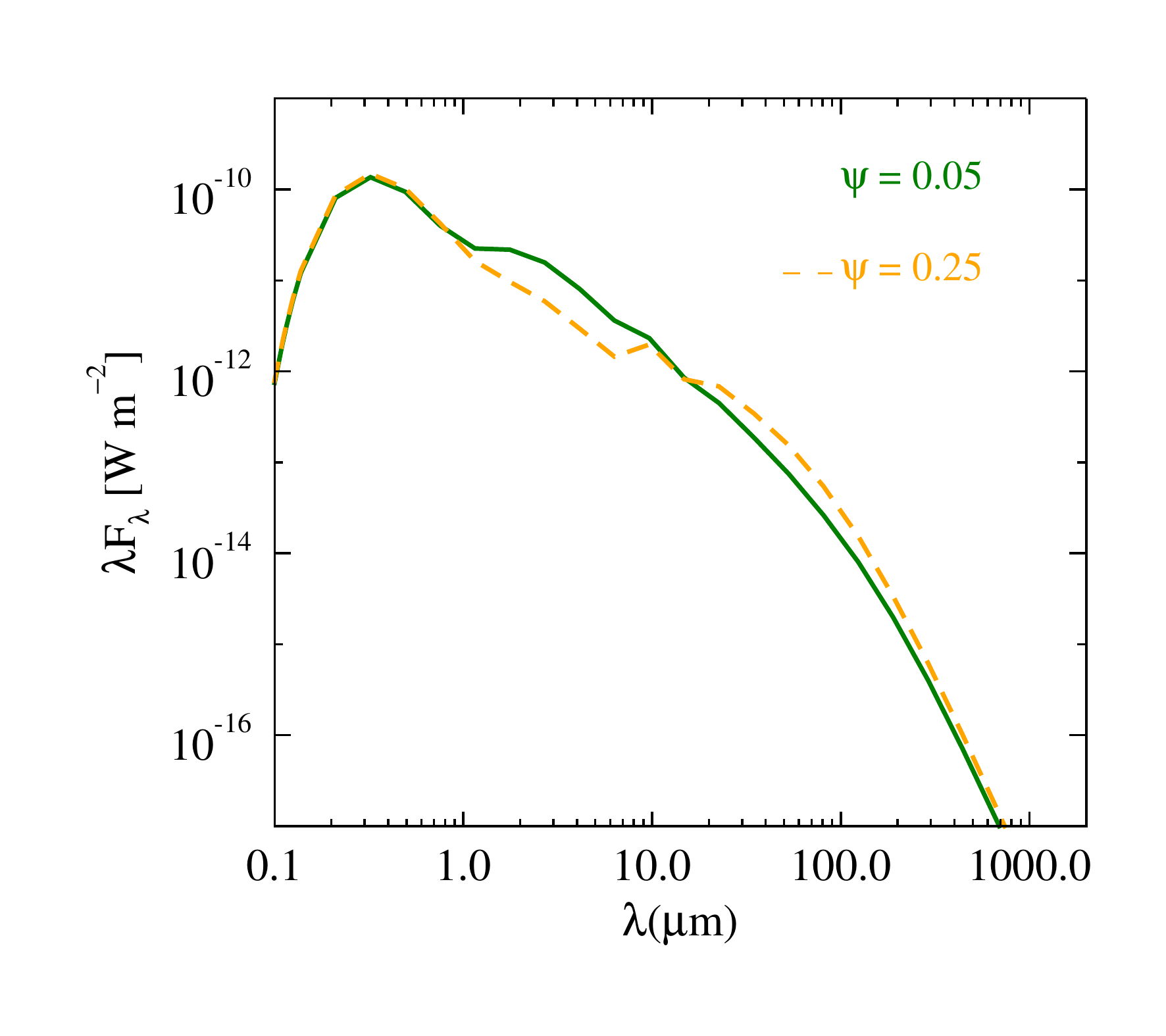}
\caption{
Spectral energy distribution of the two representative disk models.
}
\label{fig:sed}
\end{figure}

The Spectral energy distribution of the two representative disk models is shown in Figure~\ref{fig:sed}. 

\smallskip
\noindent

\end{appendix}

\end{document}